# Multi-Criteria Decision Making in Chemical and Process Engineering: Methods, Progress and Potential

**Zhiyuan Wang [a], Seyed Reza Nabavi [b], Gade Pandu Rangaiah [c, d, *]**

[a] Department of Computer Science, DigiPen Institute of Technology Singapore, Singapore 828608, Singapore

[b] Department of Applied Chemistry, Faculty of Chemistry, University of Mazandaran, Babolsar, Iran

[c] Department of Chemical and Biomolecular Engineering, National University of Singapore, Singapore 117585, Singapore

[d] School of Chemical Engineering, Vellore Institute of Technology, Vellore 632014, India

* Corresponding author: Gade Pandu Rangaiah, Email: chegpr@nus.edu.sg

## Abstract

Multi-criteria decision making (MCDM) is necessary for choosing one from the available alternatives (or from the Pareto-optimal solutions obtained by multi-objective optimization), where the performance of each alternative is quantified against several criteria (or objectives). This paper presents a comprehensive review of the application of MCDM methods in chemical and process engineering. It systematically outlines the essential steps in MCDM, including the various normalization, weighting, and MCDM methods that are critical to the decision-making. The review draws on published papers identified through a search in the SCOPUS database, focusing on works by authors with more contributions to the field and on highly cited papers. Each selected paper is analyzed based on the MCDM, normalization and weighting methods used. Additionally, this paper introduces two readily available programs for performing MCDM calculations. In short, it provides insights into MCDM steps and methods, highlights their applications in chemical engineering, and discusses the challenges and prospects in this area.

## 1. Introduction

Optimization is an important area within chemical engineering, and it has found numerous applications in chemical engineering, like in many other disciplines. Traditionally, optimization is for minimizing or maximizing a single objective (i.e., single objective optimization, SOO). In the last 25 years, optimization for multiple objectives (i.e., multi-objective optimization, MOO) has attracted increasing interest from chemical and process engineers and has now become common for applications. See the review by Cui et al. (2017) on MOO in the energy area, Madoumier et al. (2019) on MOO in food processing, and Rangaiah et al. (2020) on MOO in chemical engineering. Both SOO and MOO problems involve one or more decision variables, each with its lower and upper bounds, and may have equality and/or inequality constraints.

SOO of an application generally gives only one optimal solution (i.e., a minimum and a maximum in case of minimization and maximization applications, respectively). On the other hand, objectives in MOO are often conflicting, which means there is no single optimal solution for simultaneously optimizing all the objectives in the application. Hence, MOO provides a set of optimal solutions, known as Pareto-optimal, non-dominated, non-inferior and efficient solutions. Objective values of any one of these solutions cannot be further improved without worsening one or more objective values of another solution in the set of Pareto-optimal solutions.

Pareto-optimal solutions for a process optimization problem having two objectives to be maximized, are illustrated in Figure 1, which shows the optimal values of objectives only. Optimal values of decision variables, although crucial in implementing an optimal solution, are not required for understanding the trade-off relationship of Pareto-optimal solutions and so they are not shown in Figure 1. This figure is for maximizing both ethylene selectivity and ethane conversion simultaneously, in an industrial reactor producing ethylene from ethane by steam cracking (Tarafder et al., 2005). Figure 1 illustrates Pareto-optimal solutions as filled blue circles, while filled red triangles represent the dominated/inferior solutions. Both the objectives (i.e., ethane conversion and ethylene selectivity) of these Pareto-optimal solutions cannot be improved simultaneously. As depicted in Figure 1, improvement in one objective (e.g., ethylene selectivity) of any Pareto-optimal solution is accompanied by worsening of the other objective (e.g., ethane conversion); in other words, there is trade-off between the two objectives. The number of Pareto-optimal solutions obtained by solving an MOO problem is usually large (say, 50 or more). The objectives in chemical engineering applications can be related to fundamentals (e.g., conversion, selectivity and recovery), economics (e.g., capital cost, operating cost, payback period and net present value), energy (e.g., steam required, electricity required and exergy loss), environment (e.g., $CO_2$ emissions and environmental impact), control (e.g., integral error and condition number), safety (e.g., inventory, inherent safety index) etc. (Rangaiah et al., 2020).

In some applications, there may be a limited number (e.g., 3 to 20) of alternatives (Wang et al., 2023b), and the performance of each of them is evaluated according to several criteria. For example, Husain et al. (2024) investigated the selection of an optimal renewable energy source in India from the four alternatives: solar, wind, hydro and biomass. Performance of these alternatives is quantified in terms of 10 criteria from C1 to C10 (Table 1). Here, C1 represents the Installed Cost, ($/kW). C2 denotes the Operating & Maintenance Cost ($/kW.y).

C3 is the Levelized Cost of Electricity ($/kWh). C4 refers to Efficiency (%). C5 is the Capacity Factor (%). C6 quantifies Greenhouse Gas Emission ($gCO_2$/kWh). C7 measures Land Requirement (m2/kW). C8 denotes Job Creation (Job-years/GWh). C9 reflects Technical Maturity, on a scale of 1 to 5. Finally, C10 captures Social Acceptance, also on a scale of 1 to 5. Husain et al. (2024) compiled the criteria values of all the four alternatives in Table 1 as follows: values of C1 to C8 from the literature, values of technical maturity (C9) based on expert opinion, and values for social acceptance (C10) from a survey questionnaire. Values of C1 to C8 for each alternative may be by the design of each renewable energy system based on technical expertise or by solving an appropriate MOO problem.

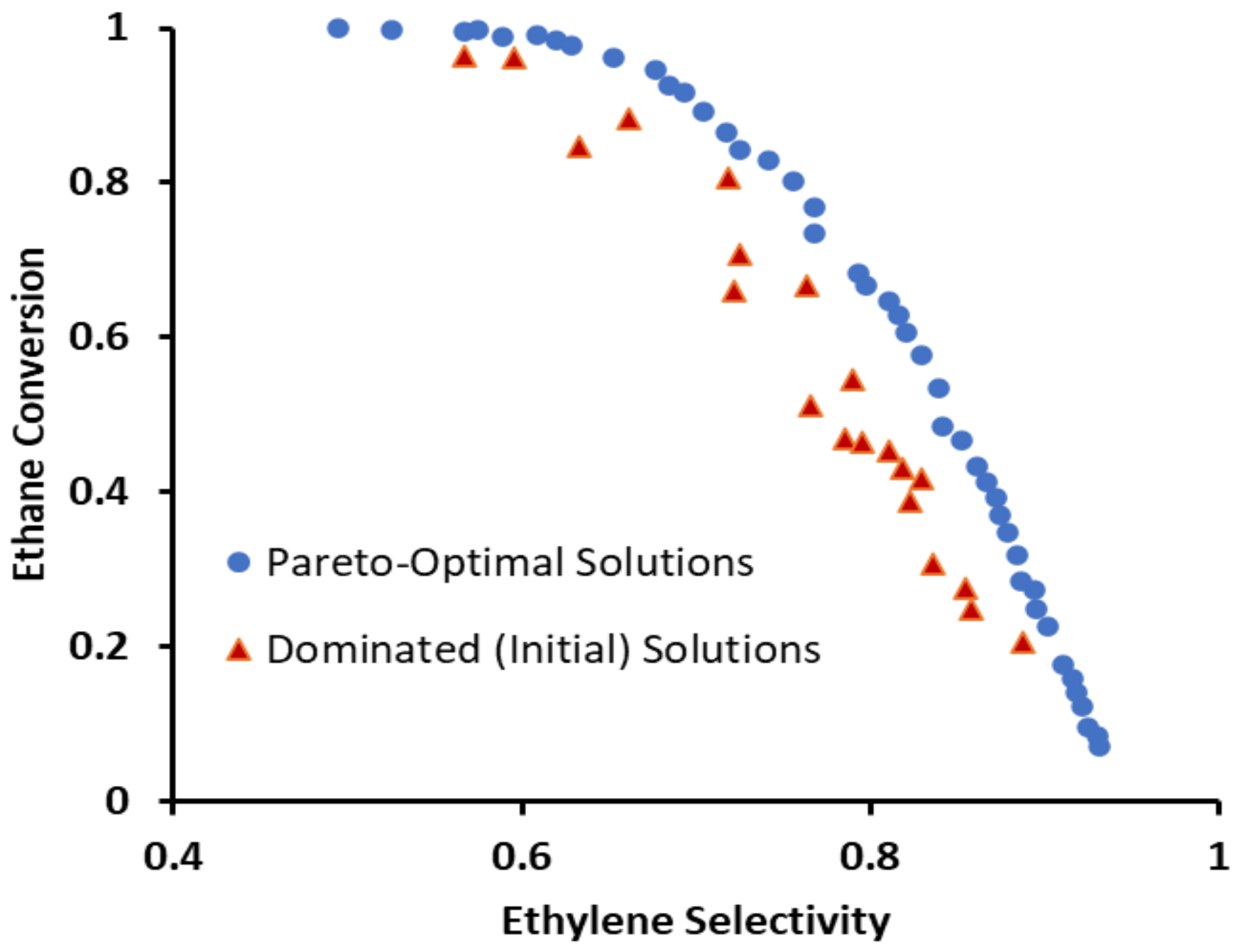

Figure 1. Dominated and Pareto-optimal solutions for simultaneous maximization of ethane conversion and ethylene selectivity of an industrial reactor producing ethylene by steam cracking of ethane.

Eventually, for any application, only one solution from the Pareto-optimal solutions (e.g., Figure 1) obtained by solving the related MOO problem or only one of the available alternatives (e.g., Table 1) is required for implementation. This selection or decision-making falls within the field of multi-criteria decision making (MCDM), which refers to decision-making in the presence of multiple, often conflicting, criteria. This analysis involves several steps, and many methods are available for them. MCDM generally ranks the given alternatives, and one of the top-ranked alternatives can be chosen for implementation.

Table 1. Alternatives and criteria for the selection of a renewable energy source in India (Husain et al., 2024). Here, criteria C1, C2, C3, C6 and C7 should be minimized whereas C4, C5, C8, C9 and C10 should be maximized.

| **Alternatives (Renewable Energy Sources)** | **Criteria** | | | | | | | | | |
|---|---|---|---|---|---|---|---|---|---|---|
| | **C1** | **C2** | **C3** | **C4** | **C5** | **C6** | **C7** | **C8** | **C9** | **C10** |
| **Solar** | 596 | 9,000 | 0.038 | 22 | 19 | 48 | 12 | 0.87 | 4 | 4.58 |
| **Wind** | 1038 | 28,000 | 0.04 | 35 | 33 | 11 | 250 | 0.17 | 4 | 4.17 |
| **Hydro** | 1817 | 45,425 | 0.065 | 76.61 | 57 | 24 | 500 | 0.27 | 5 | 3.56 |
| **Biomass** | 1154 | 46,160 | 0.057 | 84.33 | 68 | 230 | 13 | 0.21 | 3 | 4.00 |

MCDM has been researched in diverse disciplines including economics, business, healthcare, transportation, and industrial engineering. Partly because of its interdisciplinary nature, the terminology in MCDM field varies. For example, criterion, attribute, and objective have the same significance; MCDM is also known as multi-attribute decision making (MADM), multi-criteria decision analysis/aid/aiding (MCDA), and multi-criteria analysis (MCA). On the other hand, MCDM problems are classified into two groups by Hwang & Yoon (1981): MADM and multi-objective decision making (MODM), of which the latter is essentially MOO outlined above and common in (chemical) engineering. For MOO, interested readers can refer to the review paper by Rangaiah et al. (2020) and the book by Rangaiah (2017), for its techniques and applications in chemical engineering. The present paper adopts the following terminology: criteria (synonymous with objectives and attributes), alternatives (synonymous with Pareto-optimal, non-dominated, non-inferior and efficient solutions), MCDM (synonymous with MADM, MCDA and MCA) and MOO (synonymous with MODM). Depending on the application, MOO may be required to find the Pareto-optimal solutions (i.e., alternatives) before MCDM.

Although MOO has become popular in chemical engineering, application of and studies on MCDM in chemical engineering field are limited. However, MCDM is essential for selecting one of the Pareto-optimal solutions found by MOO. Moreover, it is required for decision-making when there are competing alternatives (for a chemical process, feedstock choices, material selections, site locations, vendors, supply chains etc.) and multiple criteria. Hence, the broad aims of this paper are to: (1) highlight the need for MCDM in chemical and process engineering, (2) outline the steps and methods for MCDM, (3) present MCDM programs, (4) review MCDM applications in chemical engineering, (5) outline studies in chemical engineering applications and (6) discuss challenges and opportunities in MCDM.

This work uniquely contributes to the emerging field by providing concise overview of MCDM and in-depth review of MCDM applications in chemical and process engineering. This is motivated by the limited application and study of MCDM in chemical and process engineering. The overview covers main steps in MCDM, normalization techniques, some weighting approaches and popular/recent MCDM methods; it offers tailored insights for researchers and practitioners new to MCDM. Further, a distinctive aspect of this paper is the exploration of practical tools, e.g., an Excel VBA-based program (i.e., EMCDM) and a Python library (i.e., PyMCDM), which are readily accessible. Then, we review some MCDM applications in chemical engineering and discuss the potential of integrating MCDM methods with emerging technologies, such as artificial intelligence (AI) and machine learning (ML) algorithms (Wang et al., 2022b), providing a future-oriented perspective on decision-making advancements. This approach not only bridges theory and practice but also establishes a foundation for future studies aimed at more complex MCDM problems in chemical and process engineering.

The rest of this paper is organized as follows. Section 2 outlines the essential steps in MCDM. Sections 3 and 4 describe the normalization and weighting methods, respectively. Section 5 reviews popular and/or recent MCDM methods. Section 6 covers two MCDM programs available to interested readers. Sections 7 and 8 explore the application of MCDM in chemical and process engineering as reported in journal papers. Section 9 discusses the challenges associated with MCDM. Finally, Section 10 presents the conclusions of this study.

## 2. Steps in Multi-Criteria Decision Making

The general steps of MCDM are depicted in Figure 2. Initially, as aforementioned, numerical solution of the related MOO problem may be required to generate a set of Pareto-optimal solutions. These solutions or alternatives represent the best possible trade-offs among conflicting criteria, without any one solution being superior across all criteria. Following this, alternatives-criteria matrix (ACM), like that in Table 1, should be meticulously constructed. This matrix includes all the alternatives, and their values of criteria considered.

After constructing the ACM, the next crucial step is the normalization of criteria values to ensure comparability by transforming them to a common scale. This can be accomplished through various normalization methods such as Vector normalization, Sum normalization, Max-Min normalization, and Max normalization. Each of these methods adjusts the criteria values differently based on its governing principle. In general, the normalized criteria values are in the range of 0 to 1. More details of normalization methods are given in Section 3.

After normalization, assigning weights to each of the criteria is often necessary, reflecting the relative importance of each criterion as perceived by decision-makers. Note that weights can be determined subjectively by decision-makers or through weighting methods such as entropy method, criteria importance through intercriteria correlation (CRITIC) method, stepwise weight assessment ratio analysis (SWARA) method, or even simple methods like equal weighting (i.e., mean method). Principles of selected weighting methods are covered in Section 4.

The culmination of the MCDM involves ranking the alternatives in the weighted, normalized ACM to identify the most preferable one(s). Various MCDM methods in the literature can be employed for this purpose (Nabavi et al., 2023a; Nabavi et al., 2024), such as the technique for order of preference by similarity (TOPSIS), simple additive weighting (SAW), or multi-attributive border approximation area comparison (MABAC). Each method has a different underlying algorithm to calculate the performance scores and determine the ranking of alternatives in the weighted, normalized ACM. Classification of MCDM methods and some MCDM methods used in chemical and process engineering are outlined in Section 5.

Finally, decision-makers can review the top-ranked alternative(s), make the (final) decision and terminate. If necessary, they can decide to perform sensitivity analysis by making changes in ACM, normalization, weighting and/or MCDM methods, and repeat MCDM steps. Sometimes, an MOO problem needs to be updated to consider potential changes in the chemical process (e.g., feed availability, product demand and feed/product prices) and solved to update the Pareto-optimal solutions. Alternatively, number of alternatives and/or criteria values may have to be adjusted. Ranking of alternatives with another normalization, weighting and/or MCDM methods can be studied as well. Sensitivity analysis provides insights into the effect of potential changes in various steps of MCDM on the ranking of all alternatives as well as top-ranked alternatives.

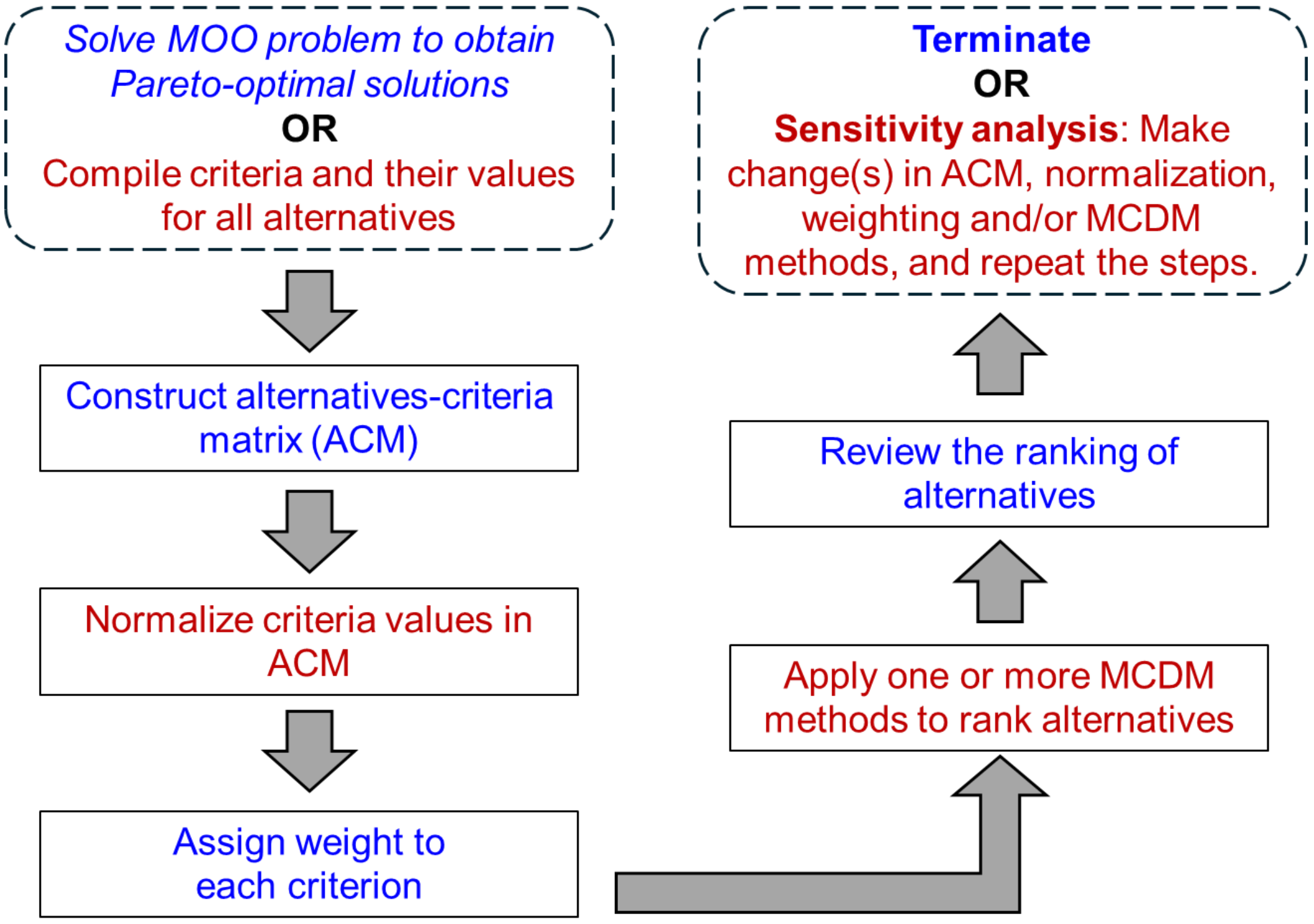

Figure 2. General steps in MCDM

## 3. Normalization Methods

The notation and symbols employed in this paper are as follows. ACM consists of $m$ rows (i.e., $m$ alternatives) and $n$ columns (i.e., $n$ criteria). A criterion can be for maximization (i.e., a benefit criterion) or minimization (i.e., a cost criterion). The symbol $f_{ij}$ denotes the value (or measurement) of the $j^{th}$ criterion for the $i^{th}$ alternative in the ACM; here, $i \in \{1,2,\ldots,m\}$ and $j \in \{1,2,\ldots,n\}$. Likewise, $f_{kj}$ is the value of the $j^{th}$ criterion for the $k^{th}$ alternative, with $k \in \{1,2,\ldots,m\}$. Additionally, the symbol $F_{ij}$ represents the normalized $f_{ij}$ value. The symbol $w_j$ indicates the weight assigned to the $j^{th}$ criterion.

Table 2 presents various normalization methods utilized to normalize the criteria values of differing magnitudes to a common scale. Each method comes with two formulas, a direct normalization formula applicable to maximization criteria, and an inverse normalization formula for minimization criteria. As the name implies, the inverse normalization formula effectively reverses the optimization direction, meaning that if the $j^{th}$ criterion ($f_{ij}$) is originally for minimization (smaller-the-better), its normalized counterpart ($F_{ij}$) would be transformed to maximization (larger-the-better) type, owing to the introduction of a negative sign to $f_{ij}$ or the use of its reciprocal.

Table 2. Normalization methods

| Normalization method | Direct normalization (for maximization criteria) $i \in \{1,2,\dots,m\}$ $k \in \{1,2,\dots,m\}$ $j \in \{1,2,\dots,n\}$ | Inverse normalization (for minimization criteria) $i \in \{1,2,\dots,m\}$ $k \in \{1,2,\dots,m\}$ $j \in \{1,2,\dots,n\}$ |
|---|---|---|
| Sum normalization (Kosareva et al., 2018) | $F_{ij} = \frac{f_{ij}}{\sum_{k=1}^{m} f_{kj}}$ | $F_{ij} = \frac{(1/f_{ij})}{\sum_{k=1}^{m}(1/f_{kj})}$ |
| Vector normalization (Kosareva et al., 2018) | $F_{ij} = \frac{f_{ij}}{\sqrt{\sum_{k=1}^{m} {f_{kj}}^2}}$ | $F_{ij} = 1 - \frac{f_{ij}}{\sqrt{\sum_{k=1}^{m} {f_{kj}}^2}}$ |
| Max-Min normalization (Kosareva et al., 2018) | $F_{ij} = \frac{f_{ij} - min_k(f_{kj})}{max_k(f_{kj}) - min_k(f_{kj})}$ | $F_{ij} = \frac{max_k(f_{kj}) - f_{ij}}{max_k(f_{kj}) - min_k(f_{kj})}$ |
| Max normalization (Kosareva et al., 2018) | $F_{ij} = \frac{f_{ij}}{max_k(f_{kj})}$ | $F_{ij} = \frac{min_k(f_{kj})}{f_{ij}}$ |
| Logarithmic normalization (Zavadskas & Turskis, 2008) | $F_{ij} = \frac{\ln(f_{ij})}{\ln(\prod_{k=1}^{m} f_{kj})}$ | $F_{ij} = \frac{1 - \frac{\ln(f_{ij})}{\ln(\prod_{k=1}^{m} f_{kj})}}{m-1}$ |
| Peldschus non-linear normalization (Zavadskas & Turskis, 2008) | $F_{ij} = \left(\frac{f_{ij}}{max_k(f_{kj})}\right)^2$ | $F_{ij} = \left(\frac{min_k(f_{kj})}{f_{ij}}\right)^3$ |

It is important to highlight that not all MCDM and weighting methods employ both direct and inverse normalization formulas in their original algorithms. For instance, both the original TOPSIS method proposed in Hwang & Yoon (1981) and the original MOORA (MOO on the basis of ratio analysis) method introduced by Brauers & Zavadskas (2006) solely employ the direct normalization formula of the vector normalization method. The entropy weighting method described in Hwang & Yoon (1981) utilizes only the direct normalization formula of the sum normalization method.

## 4. Weighting Methods

To assign weight to each criterion, decision-makers may leverage their domain expertise for weight determination. Alternatively, they can employ established weighting methods, which are divided into two categories: subjective and objective weighting methods. For subjective weighting methods, decision-makers are required to provide their preference assessments with respect to the criteria, thereby reflecting their personal judgments and insights. On the contrary, objective weighting methods do not require such preference assessments from decision-makers; instead, these methods derive weights solely based on the constructed

ACM. Some popular objective and subjective weighting methods, shown in Figure 3, are outlined in the following two subsections.

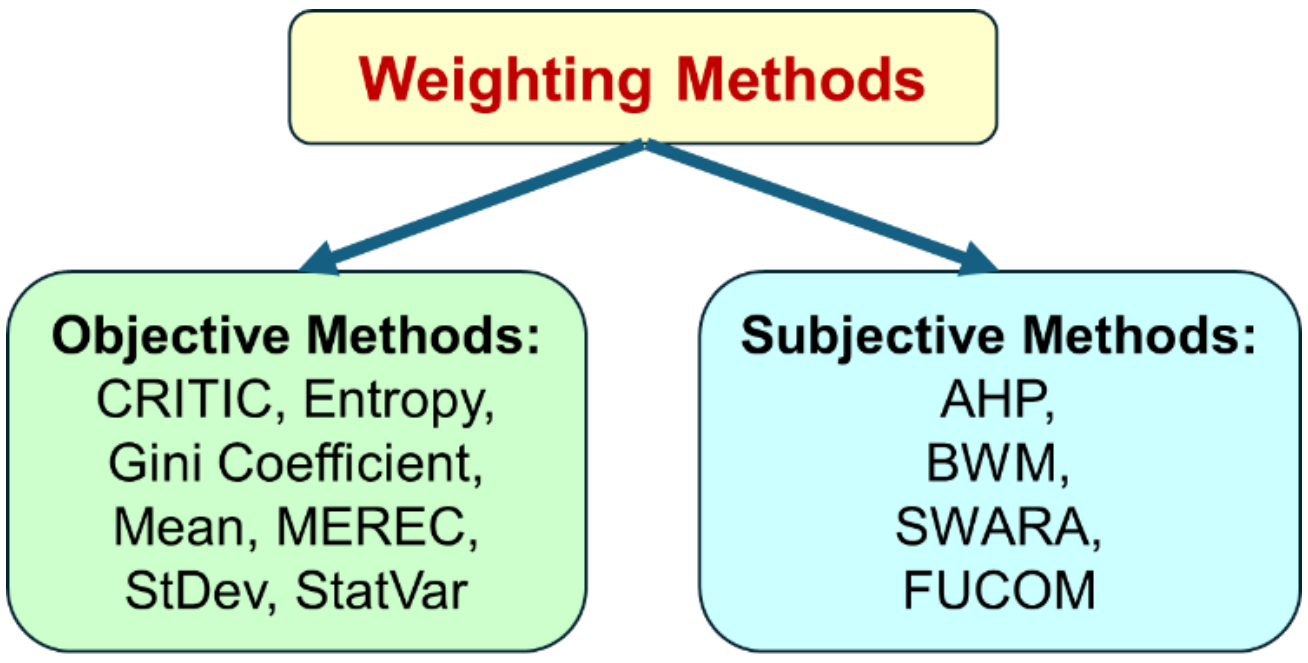

Figure 3. Objective and subjective weighting methods described in Sections 4.1 and 4.2 respectively.

### 4.1. Objective Weighting Methods

**Criteria importance through intercriteria correlation (CRITIC)** method evaluates the pairwise correlations among the criteria within the ACM along with the standard deviation of each criterion, to determine their weights. As delineated in Diakoulaki et al. (1995) and Vujičić et al. (2017), the evaluation begins by normalizing the original ACM using the max-min normalization; then, pairwise correlations between all criteria are calculated to form a symmetric $n \times n$ correlation matrix, where each value ranges from –1 to +1; –1 indicates a completely negative correlation, +1 indicates an entirely positive correlation, and 0 indicates no correlation between the two criteria. Following this, the standard deviation of values of each criterion is computed. Finally, weights are derived by integrating the correlation values and the standard deviations, following the procedural equations in Diakoulaki et al. (1995). The CRITIC method captures the diversity of criterion values and their interdependencies, providing a comprehensive approach to weight assignment. See Diakoulaki et al. (1995) for the steps and equations of this method.

**Entropy** weighting method relies on a measure of uncertainty in the information, expressed through the principles of probability theory. When there is substantial variation in the values of a particular criterion across the alternatives in ACM, this criterion receives a correspondingly higher weight. The entropy weighting method starts with normalizing the original ACM using the sum normalization, as described in Hwang & Yoon (1981) and many other studies (Mon et al., 1994; Zardari et al., 2014; Dos Santos et al., 2019). Next, the aggregated entropy value for each criterion is computed. The criteria weights are then derived based on these entropy values, effectively translating the degree of uncertainty into a quantitative weighting metric. See Sałabun et al. (2020) for the steps and equations of entropy weighting method.

**Gini coefficient** weighting method, introduced in Li & Chi (2009), applies the concept of the Gini coefficient, traditionally utilized to measure income inequality (Öztaş et al., 2023). It capitalizes on the variability among criteria values within ACM to assign weights, considering the differences in data distribution across criteria. It does not require an explicit normalization of the original ACM. The Gini coefficient for each criterion is calculated by assessing the differences between pairs of all alternatives for that criterion, adjusted by the average value of that criterion (somewhat like normalization). Next, weight of each criterion is determined by dividing its Gini coefficient by the sum of Gini coefficients of all criteria, which ensures that the sum of all weights equals one. See Li & Chi (2009) for the steps and equations of Gini coefficient method.

**Mean** weighting method is the simplest approach to allocating weights to criteria. It is based on the assumption that all criteria are of equal importance, thereby allocating an equal weight to each criterion (Zardari et al., 2014). This method does not require any additional calculations or normalization, making it straightforward and applicable to cases where a uniform assessment of criteria is deemed appropriate or as an initial trial.

**Method based on the removal effects of criteria (MEREC)** is a recent weighting method introduced by Keshavarz-Ghorabaee et al. (2021). It determines criteria weights based on the removal effect of each criterion on the overall performance of alternatives. A criterion is assigned a higher weight if its removal significantly affects the aggregate performance assessment. The process begins with the normalization of the ACM using max normalization; notably, MEREC employs the direct normalization formula in the max normalization method for minimization objectives, as well as the inverse normalization formula for maximization objectives, which is opposite to the conventional practices (Table 2). Then, considering all criteria, the overall performance assessments of all alternatives are calculated, followed by the performance assessments with each criterion removed one by one. Next, for each criterion, the aggregated absolute deviation is computed based on these performance assessments. The weight of each criterion is then calculated by dividing its aggregated absolute deviation by the sum of the aggregated absolute deviations for all criteria. See Keshavarz-Ghorabaee et al. (2021) for the steps and equations of MEREC method.

**Standard deviation (StDev)** weighting method assigns weights to criteria based on their standard deviation values. It begins by normalizing the original ACM using max-min normalization as outlined in Diakoulaki et al. (1995) and Mukhametzyanov (2021). Next, standard deviation of each criterion is calculated. The weight of each criterion is then determined by dividing its standard deviation by the sum of the standard deviations of all criteria. See Diakoulaki et al. (1995) for the steps and equations of StDev method.

**Statistical variance (StatVar)** weighting method closely resembles the StDev method, with the primary distinction being its use of variance values of criteria to determine the weights. Essentially, both StDev and StatVar methods directly operate on the principle that if the values of a criterion across all alternatives exhibit a small variance (say, in the extreme case, where all values of this criterion are identical, resulting in a variance of zero), a small weight should be assigned to that criterion. This is because such a criterion does not help much in distinguishing the alternatives (Shuai et al., 2012). See Rao & Patel (2010) for the steps and equations of StatVar method.

### 4.2. Subjective Weighting Methods

**Analytic hierarchy process (AHP)** method, originally developed by Saaty (1990), is one of the widely applied methods for determining criteria weights based on decision-makers' subjective pairwise comparisons. To facilitate this, Saaty's 1-9 assessment scale (Table 3) is typically employed. The AHP method begins with constructing a decision hierarchy, organizing the decision goal (e.g., decide on the optimal renewable energy source in India), criteria (e.g., installed cost, efficiency, greenhouse gas emission, land requirement etc.) and alternatives (e.g., solar, wind, hydro, etc.) into a structured model. Decision-makers then perform subjective pairwise comparisons, evaluating the relative importance of each pair of criteria using the Saaty's scale. These comparisons are compiled into a $n \times n$ pairwise comparison matrix. Following this, the eigenvalue method is used to find the priority eigenvector from these matrices, which then determines the weights of the criteria. Instead of this method, there is a simpler approximation technique (using sum normalization), which is popular in many applications (Teknomo, 2006; Karim & Karmaker, 2016; Wang et al., 2020; Khan & Nazir, 2023).

In addition to determining criteria weights, it is crucial to recognize that the AHP method can be further utilized for decision making directly (i.e., act as a MCDM method). For this, AHP requires decision-makers to conduct subjective pairwise comparisons among all pairs of alternatives with respect to each criterion, leading to $n$ number of $m \times m$ pairwise comparison matrices (these are different from the pairwise comparison matrix based on criteria, used for weights in the previous paragraph). According to the original AHP method in Saaty (1990), all these pairwise comparisons are purely subjective, sourcing directly from decision-makers. The typical ACM, like that in Table 1, is not immediately compatible with the AHP acting as a MCDM method. However, mapping of the objective quantitative data to Saaty's 1-9 scale, like what Si et al. (2016) performed, is possible. Further discussion on AHP as a MCDM method is presented in Section 5. In the present subsection, the focus is on describing AHP as a subjective weighting method. See Saaty (1990) for the steps and equations of AHP method.

Table 3. Saaty's 1-9 pairwise comparison scale

| Scale | Compare between events *x* and *y* |
|---|---|
| 1 | Equal importance of *x* against *y* |
| 3 | Moderate importance of *x* against *y* |
| 5 | Essential or strong importance of *x* against *y* |
| 7 | Very strong importance of *x* against *y* |
| 9 | Extreme importance of *x* against *y* |
| 2, 4, 6, 8 | Intermediate value between adjacent scales |
| Any decimal within [1, 9], e.g., 2.3 and 5.1 | Finer scale |
| Reciprocals, e.g., 1/3 and 1/7 | If event *x* has one of the above numbers assigned to it when compared with event *y*, then *y* has the reciprocal value when compared with *x*. |

**Best-worst method (BWM**), originally developed by Rezaei (2015), employs two vectors of pairwise comparisons to determine the weights of criteria. It yields multiple sets of weights. Building on this concept, Rezaei (2016) proposed a linear approach that generates unique set of weights. This method distinguishes itself by identifying the most and least preferred criteria (i.e., best and worst criteria that decision-makers specify) and then using these benchmarks to assess the relative importance of all other criteria. BWM utilizes two vectors: the best-to-others (BO) vector, where the best criterion is compared to all others using a scale from 1 to 9, and the others-to-worst (OW) vector, which compares all criteria against the worst criterion. The weights of the criteria are then determined through a linear optimization model that seeks to minimize the inconsistency among the comparisons. BWM is simple and efficient in deriving consistent criteria weights. See Rezaei (2016) for the steps and equations of BWM method.

**Step-wise weight assessment ratio analysis (SWARA)** method, developed by Keršulienė et al. (2010), is another subjective weighting method. Like AHP, it derives weights based on decision-makers' preferences. SWARA begins by arranging the criteria in descending order of importance as determined by decision-makers. Following this initial sorting, decision-makers need to assign a relative importance value, ranging from 0 to 1, for each criterion relative to its predecessor, starting from the second criterion. The subsequent calculations involve deriving the coefficients $K_j$ and $Q_j$ using specific formulas (Keršuliene et al., 2010). The final weight for each objective is then determined by normalizing $Q_j$ across all objectives.

**Full Consistency Method (FUCOM)**, developed by Pamučar et al. (2018), is a novel subjective weighting approach that derives criteria weights based on pairwise comparisons with an emphasis on achieving full consistency in decision-makers' preferences. It starts by having decision-makers establish the relative importance of criteria through pairwise comparisons, arranging them in descending order of importance. Next, decision-makers specify preference ratios for each pair of criteria, comparing each criterion's importance relative to the next. The consistency of these pairwise comparisons is then checked using a consistency coefficient. If necessary, adjustments are made to ensure the final weight values satisfy the full consistency condition.

Other novel approaches for subjective weighting are RANCOM (Więckowski et al., 2023) and KEMIRA (Krylovas et al. 2014).

### 5. MCDM Methods

There are more than 200 MCDM methods, which can be classified into several types, as shown in Figure 4. This classification is based on the grouping in Ishizaka & Nemery (2013) and Thakkar (2021). The commonly used methods in each type are also shown in Figure 4. They are outlined in the following subsections, wherein the methods are covered in alphabetical order. Detailed steps and equations of each method are available in the respective cited reference. Criteria weights are typically required for MCDM methods presented here; the exception is the GRA variant described in Subsection 5.1, which does not require weights.

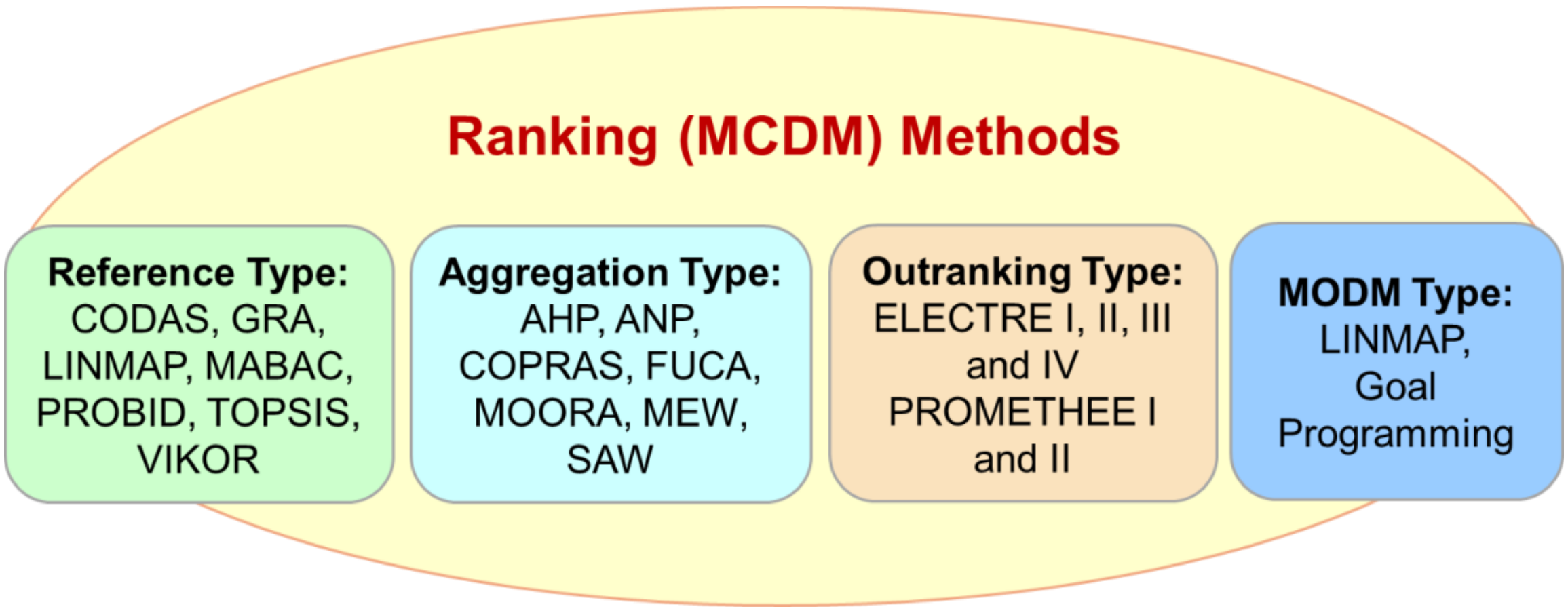

Figure 4. Classification of MCDM methods, and popular Reference, Aggregation, Outranking and MODM type MCDM Methods

#### 5.1. Reference Type MCDM Methods

The reference type MCDM methods are a class of decision-making methods that evaluate and rank alternatives by measuring their performance relative to one or more ideal reference solutions. These ideal solutions encompass the positive ideal solution having the best values of each criterion in the ACM, and the negative ideal solution having the worst values of each criterion in the ACM.

**Combinative distance-based assessment (CODAS)** method of Keshavarz-Ghorabaee (2016), is underpinned by the core principle of finding the performance score for each alternative by calculating both Euclidean and Taxicab distances from the negative ideal solution. Inherently, an alternative that exhibits greater distance from this negative ideal solution is considered more favorable. As stated in Keshavarz-Ghorabaee (2016), if there are two alternatives that are incomparable according to Euclidean distance, then Taxicab distance is used as the secondary measure.

**Gray/Grey relational analysis (GRA)** method, based on the principles of gray system theory (Deng, 1982), has multiple variants in the MCDM literature. The specific variant we reference, as outlined by Martinez-Morales et al. (2010), is notable by its autonomy from user-specified inputs (e.g., weight for each criterion). A fundamental step in GRA method involves finding the positive ideal solution. Next, the method computes the distances between each alternative and this ideal solution to compute the gray relational coefficient. This coefficient assesses the similarity of each alternative to the ideal reference, enabling the ranking of alternatives.

**Linear programming technique for multidimensional analysis of preference (LINMAP)**, developed by Srinivasan & Shocker (1973), is another method for ranking the alternatives in the ACM. Like other reference type MCDM methods, a positive ideal solution needs to be identified. Next, the method involves calculation of distances between each alternative and this ideal solution. Finally, the alternative that lies closest to the positive ideal solution is recommended. This procedure presumes that decision-makers prefer the alternative closest to the positive ideal solution, thus facilitating a direct and intuitive assessment of options based on their proximity to this ideal. Note that LINMAP can also be employed to estimate the criteria weights using the linear programming technique. See Srinivasan & Shocker (1973) for the detailed steps and equations of LINMAP method, including linear programming formulation used to calculate weights. Alternatively, refer to Sayyaadi & Mehrabipour (2012) for the calculation of distances and ranking for alternatives.

**Multi-attributive border approximation area comparison (MABAC)** method of Pamučar & Ćirović (2015), focuses on calculating the sum of the distances of alternatives from a unique reference solution known as the border approximation area matrix. This matrix is derived from the product of the weighted normalized values of each criterion across all alternatives. Like other MCDM methods, MABAC starts with the construction of a weighted normalized ACM from the original ACM. This is followed by determining the border approximation area value for each criterion, collectively forming the border approximation area matrix. Next, the distances between each alternative and the border approximation area matrix are computed. These distances are used to derive assessment scores and rankings for the alternatives.

**Preference ranking on the basis of ideal-average distance (PROBID)** method, developed by Wang et al. (2021), distinguishes itself by evaluating alternatives against a spectrum of reference solutions. This spectrum encompasses a hierarchy of ideal solutions, starting from the most positive ideal solution and extending through lesser tiers (e.g., 2nd and 3rd most positive ideal solutions) down to the most negative ideal solution. Further, PROBID method incorporates an average solution to provide a comprehensive baseline for comparison. Its procedure involves calculating the distances between each alternative and the different tiers of ideal solutions, and uniquely integrating these distances in an inversely weighted manner (e.g., coefficients 1, 1/2, and 1/3 are multiplied with the distances to the most positive ideal, 2nd positive ideal, and 3rd positive ideal solutions, respectively). This integration, combined with the distance from the average solution, is used to compute an overall performance score.

**Technique for order of preference by similarity to ideal solution (TOPSIS) method**, formulated by Hwang & Yoon (1981), stands as one of the most prominent reference type MCDM methods, employed extensively across a variety of disciplines. It requires the identification of two reference solutions, that is, the positive ideal and the negative ideal solutions. TOPSIS evaluates alternatives based on their distances from these reference points. The top-ranked alternative is that having the smallest distance to the positive ideal solution and simultaneously the largest distance to the negative ideal solution. This dual assessment in TOPSIS provides a more balanced evaluation of the alternatives.

**Visekriterijumska optimizacija i kompromisno resenje (VIKOR)**, a phrase of Serbian origin, standing for Multi-criteria Optimization and Compromise Solution, is another commonly used reference type MCDM method. It was originally introduced as one applicable technique by Opricovic (1998), anchored in the principle of identifying a compromise solution. Like TOPSIS, it needs two reference solutions: the positive ideal and the negative ideal solutions. These are utilized to compute the utility and regret measures for each alternative, where utility assesses how closely an alternative approaches the positive ideal solution, while regret measures the extent of deviation from the negative ideal solution. Alternatives are then ranked based on a composite index that integrates these two measures, ensuring a balanced evaluation. As described in Opricovic & Tzeng (2004), if alternatives are close in terms of the utility measure, the regret measure serves as a secondary criterion. See Opricovic & Tzeng (2004) for the steps and equations of VIKOR method.

### 5.2. Aggregation Type MCDM Methods

The aggregation type MCDM methods are a class of decision-making methods that evaluate and rank alternatives by aggregating their weighted performance scores (often through additive or multiplicative operations) across multiple criteria.

**Analytic hierarchy process (AHP)**, mentioned earlier in Section 4, can be extended beyond determining the weights of *n* criteria, to function as a MCDM method for ranking alternatives. After constructing the hierarchy of overall decision goal, criteria, and alternatives, it requires decision-makers to employ Saaty's 1-9 scale and provide subjective pairwise comparisons among all pairs of *m* alternatives with respect to each criterion, effectively leading to *n* number of *m x m* pairwise comparison matrices. As per the original AHP method described in Saaty (1990), all these pairwise comparisons are entirely subjective, sourcing directly from the judgments of decision-makers. For each *m x m* pairwise comparison matrix, AHP calculates the local priority value of each alternative with respect to the corresponding criterion using a technique like that utilized for determining criteria weights (e.g., eigenvalue method or its approximation technique); this results in a local priority vector (sized *m x 1*) for each criterion. Subsequently, AHP iterates through all *n* criteria, calculates, and amalgamates these *m x 1* priority vectors into an *m x n* local priority matrix, in which each row indicates the priority value of an alternative with respect to all *n* criteria. This *m x n* local priority matrix is then multiplied by the *n x 1* weights vector to aggregate into an *m x 1* vector representing the global priorities of the *m* alternatives. The alternative with the highest global priority value is ranked as the top choice. Note that the ACM with quantitative and objective (i.e., not subjective judgements from decision-makers) data or measurements, e.g., the one in Table 1, is not necessary for MCDM by AHP. However, instead of subjective values for pairwise comparison matrices, the ACM can be transformed to Saaty's 1-9 scale, as demonstrated in Si et al. (2016) for the selection of green technologies for retrofitting existing buildings considering three criteria: energy saving, investment cost and payback period.

**Analytic network process (ANP)**, developed by Saaty (2005) as an extension of AHP, addresses complex decision-making situations where the elements of the problem are interdependent and cannot be structured hierarchically. Unlike AHP, which organizes factors in a strict hierarchy, ANP utilizes a network structure that allows for multiple levels of interactions between decision layers and criteria. This method involves building a network of criteria and sub-criteria that not only influence one another but also contribute to the overall decision goal. Decision-makers use pairwise comparisons not only to determine weights of criteria but also to assess the influence of different elements on each other within the network. These comparisons are then used to aggregate the influences and priorities across the network to aid decision-making.

**Complex proportional assessment (COPRAS)** method, proposed by Zavadskas et al. (1994), is another aggregation type MCDM method. It employs a systematic approach to rank alternatives based on their relative importance. After constructing the weighted normalized ACM, COPRAS calculates the sum of weighted normalized value for all benefit criteria and the sum of weighted normalized value for all cost criteria, for each alternative. Then, the two sums from each alternative are aggregated to determine the relative importance of each alternative according to specific equations of the method (Zavadskas et al., 1994). The highest scoring alternative is the top-ranked choice.

**Faire Un Choix Adéquat (FUCA)** method, described in Fernando et al. (2011), is a simple aggregation type MCDM method. It operates by initially assigning ranks to alternatives for each criterion. Specifically, with respect to one criterion, the best alternative receives rank *1*, and the worst alternative receives rank *m*. Depending on whether the criterion is for maximization or minimization, the best value refers to the largest or smallest, respectively. After ranking alternatives for all criteria, FUCA performs aggregation by computing a weighted sum of these ranks for each alternative. The alternative with the smallest aggregated rank score is recommended as the top choice.

**MOO on the basis of ratio analysis (MOORA)** method, introduced by Brauers & Zavadskas (2006), is extensively applied in various fields for MCDM. After constructing the weighted normalized ACM, the performance score for each alternative is calculated by subtracting the aggregate of cost criteria values from the aggregate of benefit criteria values. The alternative having the highest performance score is the top-ranked choice.

**Multiplicative exponent weighting (MEW)** method, also known as **weighted product method or model (WPM)**, described in Miller & Starr (1969), is another aggregation type MCDM method. After constructing the normalized ACM, typically by max normalization, MEW raises the normalized criteria values to the power of their corresponding weights (Triantaphyllou & Mann, 1989). Then, for each alternative, this method aggregates these exponentiated criteria values by multiplying them together. The alternative that yields the highest product (i.e., performance score) from this multiplication is deemed the top-ranked choice. See Wang & Rangaiah (2017) for the detailed steps and equations of SAW method.

**Simple additive weighting (SAW)** method, also known as **weighted sum method or model (WSM)**, described in Fishburn (1967) and MacCrimmon (1968), is probably the simplest aggregation type MCDM method. After constructing the weighted normalized ACM, typically by max normalization, this method sums up the weighted normalized criteria values for each alternative. The alternative with the highest sum (i.e., performance score) is the top-ranked choice. See Wang & Rangaiah (2017) for the steps and equations of SAW method. There is also a weighted aggregated sum product assessment (WASPAS) method proposed by Zavadskas et al. (2012); it utilizes a weighted sum of the performance scores from both SAW and MEW (Chakraborty et al., 2015).

### 5.3. Outranking Type Methods

The outranking type MCDM methods are a class of methods that evaluate and rank the alternatives based on pairwise comparisons across all criteria. They establish an outranking relationship by determining the degree to which one alternative is preferred over another.

**ELECTRE (elimination et choix traduisant la realité, translated to: elimination and choice expressing the reality)** family of methods are originally proposed by Roy (1991). They are distinguished by their use of outranking relations, which establish whether one alternative is preferable to another based on a series of pairwise comparisons across all criteria. These methods assess the strengths and weaknesses of each alternative, applying thresholds of significance and veto to manage the contradictions inherent in decision-making. It involves constructing a preference structure among alternatives by considering both the strength of evidence that one alternative outranks another and the intensity of any veto that overrides a suggested preference. ELECTRE methods have been developed and refined over a few decades, with several variants like ELECTRE I, II, III, and IV, as well as different extensions. See Roy and Vanderpooten (1996) for the steps and equations of ELECTRE family methods.

**PROMETHEE (preference ranking organization method for enrichment evaluations)**, developed by Brans et al. (1986), is another outranking type MCDM method. It begins with the calculation of preference indices for each criterion, which quantify the degree of preference of one alternative over another. These indices are generated using predefined preference functions that can be customized to reflect specific decision-making contexts, including the importance of certain thresholds that signify substantial differences in criteria values. The calculated preference indices for each pair of alternatives are then employed to establish a global outranking score, which serves to construct a comprehensive preference structure across all alternatives. This structure effectively ranks the alternatives from the most to the least preferred, based on their global outranking scores. Like the ELECTRE methods, PROMETHEE also has several variants, such as PROMETHEE I and PROMETHEE II, which provide partial and complete ranking solutions, respectively. See Brans and De Smet (2016) for the steps and equations of PROMETHEE.

### 5.4. MODM Type Methods

As mentioned in Section 1, some literature categorizes MODM as a subset of MCDM (Hwang & Yoon, 1981; Thakkar, 2021). MODM methods focus on finding optimal solutions for optimization problems having two or more objectives. Essentially synonymous with MOO, they aim to identify solutions that best satisfy all objectives simultaneously.

In the context of chemical engineering, commonly used MOO methodologies, such as non-dominated sorting genetic algorithm II (NSGA-II) and multi-objective particle swarm optimization (MOPSO), are less frequently applied directly within the MCDM framework itself. Instead, they are typically utilized to generate a set of non-dominated solutions prior to MCDM. However, linear programming and its extension, goal programming, prominently feature in MCDM literature. For example, LINMAP method described in Subsection 5.1 can utilize linear programming to determine criteria weights, followed by steps for ranking alternatives and decision-making.

Goal programming is a technique within the field of MCDM (Jones & Tamiz, 2016). It expands upon linear programming, enabling decision-makers to define and prioritize specific goals, each with its own target value. The method focuses on minimizing the deviations from these target values, effectively handling trade-offs among competing objectives by assigning priority levels and weights to each goal. This prioritization helps in identifying a solution that best aligns with the preferences of decision-makers and strategic objectives. Commonly seen in MCDM literature, goal programming provides a structured framework for making complex decisions in a systematic and quantified manner. See Jones & Tamiz (2016) for the recent developments in goal programming.

## 6. MCDM Programs

There is a need for robust programs that ensure accurate and consistent calculations in MCDM. While some tools have one or two MCDM methods implemented, the availability of comprehensive programs that integrate multiple methods is very limited. Such programs are invaluable for conducting thorough sensitivity analyses and comparing the outcomes of various MCDM methods. In this section, we introduce two notable programs that address this need: EMCDM, an Excel VBA-based program, and PyMCDM, a Python-based library. Both these programs, described in the following sub-sections, facilitate the application of multiple MCDM methods, offering flexibility and depth for researchers and practitioners alike.

### 6.1. EMCDM (Excel VBA-based MCDM) Program

Our group has developed a comprehensive computer program for MCDM, named EMCDM, which is based on Microsoft Excel VBA (Wang et al., 2023a). We chose Excel VBA because of the popularity of MS Excel in both academia and industry. EMCDM has already been openly shared to many interested researchers, practitioners, and students. To utilize this program, decision-makers simply need to follow the five steps shown in Figure 5. The “Overview” sheet provides a brief guide on how to use the program, and the "Results from MOO" sheet is where decision makers input their ACM in the first step. Next, in the second step, decision-makers are required to provide essential information, such as number of objectives (or criteria), number of decision variables, number of constraints, and number of solutions (or alternatives). Note that the number of decision variables and number of constraints are optional and can be given as zero if not applicable. In the third step, decision-makers are to choose the type of criteria, where "Max" denotes maximization (or benefit), and "Min" indicates minimization (or cost). The fourth step involves entering weights, which can be manually entered by decision-makers (ensuring the sum of the weights equals one) or computed using one of the eight weighting methods displayed on the right side of the main interface (Figure 5). Finally, decision-makers can click an individual MCDM method they wish to apply (e.g., FUCA) from the left side of the main interface, to obtain its calculation results; or click the "Run All Methods" button to obtain results from all 16 MCDM methods implemented in EMCDM.

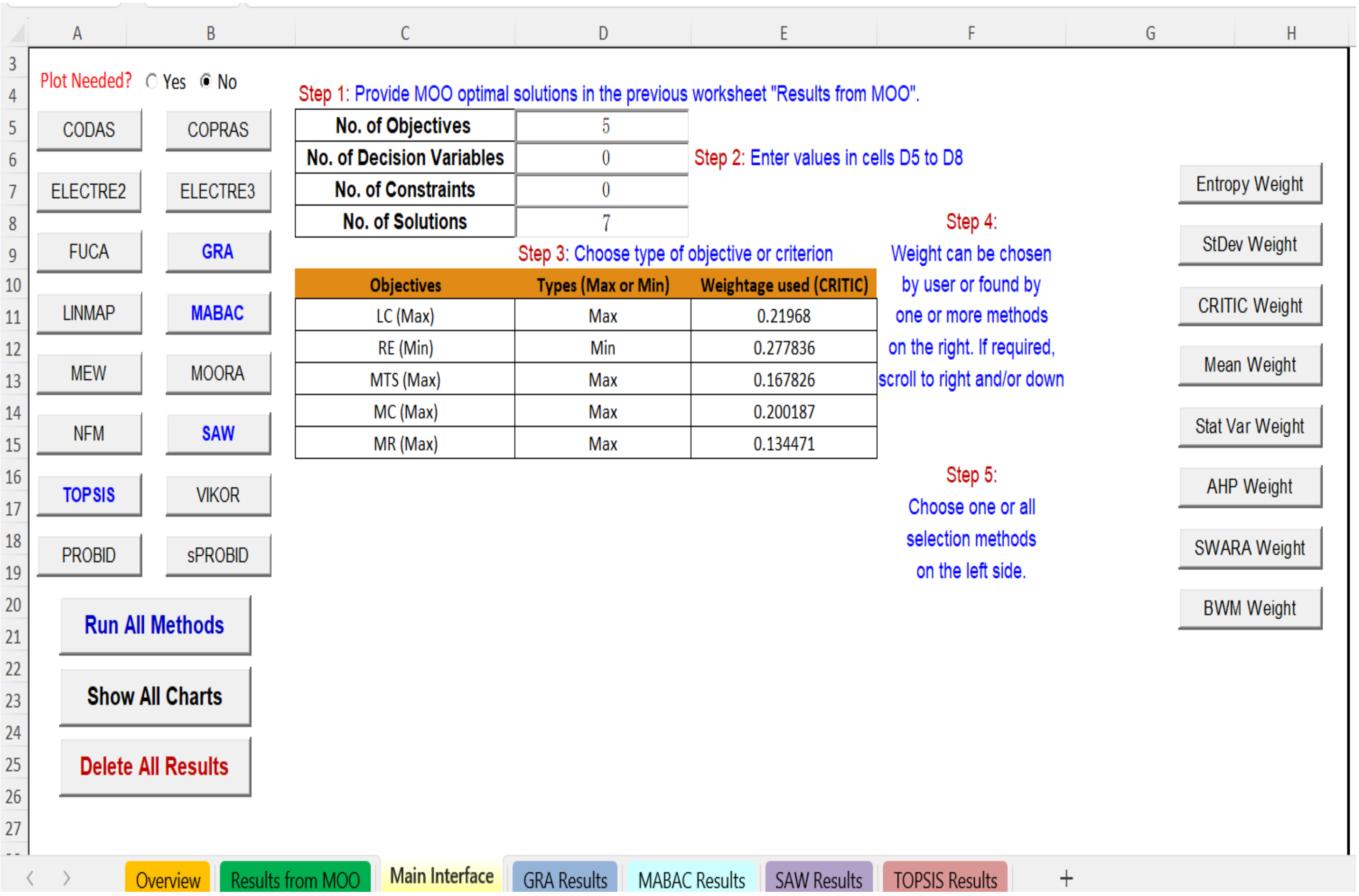

Figure 5. Main interface of the EMCDM program (version 3) with 16 MCDM methods and 8 weighting methods; the five steps for using this program are in red color.

### 6.2. PyMCDM (Python-based MCDM) Program

Recently, Kizielewicz et al. (2023) have developed a comprehensive Python-based MCDM program/library, named PyMCDM. This library leverages the capabilities of several well-known Python packages. It primarily utilizes the Numpy library for mathematical calculations and data representation through ndarray objects, and Matplotlib for data visualization. Additionally, it incorporates elements of the SciPy libraries. PyMCDM (version 1.1.0), as presented in Kizielewicz et al. (2023), has 15 MCDM methods (e.g., CODAS and TOPSIS), ten weighting methods (e.g., Entropy and CRITIC), and eight normalization methods (e.g., Sum and Vector). Users of this library need a basic understanding of Python programming and have their Python environment set up properly. For those interested in exploring this tool, Kizielewicz et al. (2023) provides sample code illustrating how to use it. Alternatively, interested readers may refer to the example provided in Table 4 as the starting point. We tried both EMCDM and PyMCDM for this example and obtained the same results for CODAS method with entropy weights.

Table 4. Sample code for using PyMCDM program

```
# !pip install pymcdm
import numpy as np
from pymcdm.methods import CODAS
from pymcdm.weights import entropy_weights
from pymcdm.helpers import rrankdata

# Define the alternatives-criteria matrix (ACM)
# 7 alternatives and 5 criteria; example from Chakraborty & Zavadskas (2014)
ACM = np.array([[60, 0.4, 2540, 500, 990],
                 [6.35, 0.15, 1016, 3000, 1041],
                 [6.8, 0.1, 1727.2, 1500, 1676],
                 [10, 0.2, 1000, 2000, 965],
                 [2.5, 0.1, 560, 500, 915],
                 [4.5, 0.08, 1016, 350, 508],
                 [3, 0.1, 1778, 1000, 920]], dtype='float')

# Find criteria weights using entropy weighting method
weights = entropy_weights(ACM)

# Define the type of each criterion, 1 for maximization, -1 for minimization
types = np.array ([1, -1, 1, 1, 1])

# Instantiate a CODAS object
codas = CODAS()

# Compute the performance scores and ranking of alternatives
performance_scores = codas(ACM, weights, types)
ranking = rrankdata(performance_scores)
```

```python
for p, r in zip(performance_scores, ranking):
    print(p, r)
```

Besides the EMCDM program and the PyMCDM library reviewed in detail here, several other MCDM tools are also available. One of them is the scikit-criteria (Cabral et al., 2016), a Python package that integrates with the broader scikit-learn environment, offers MCDM methods such as TOPSIS, VIKOR, and ELECTRE, and making it compatible with other data science and machine learning workflows. Another Python library is pyDecision (Pereira et al., 2024) that provides multiple weighting schemes and a variety of MCDM methods, including SAW, WASPAS, and MOORA.

**7. MCDM Studies in Chemical Engineering**

To find the reported studies on MCDM and its use in chemical engineering, the SCOPUS database was chosen and searched using several search terms. Search settings are summarized in Table 5. In the SCOPUS, the subject area of chemical engineering has numerous journals; of these, some journals are broad covering all engineering disciplines or science (including chemistry). Such broad journals are manually excluded to focus on those journals closely related to chemical engineering. After this screening, the number of journals included in the search is still large at more than 160. In other words, the search for MCDM studies in chemical engineering is sufficiently comprehensive and covers almost all journals of interest to chemical engineers. Here, MCDM studies refer to those articles that apply and/or analyze MCDM procedure and/or methods. Likewise, MOO studies refer to the articles that apply and/or analyze MOO.

Table 5: Settings for Search in the SCOPUS database

| | |
|---|---|
| Search Fields | Title, Abstract, Keywords |
| Publication Years | 2000 to 2023 |
| Subject Area | Chemical Engineering (see text for some journals excluded) |
| Document Type | Journal article |
| Language | English |
| Search Terms for MCDM studies | MCDM, MADM, MCDA or MCA with each of these as full form with/without hyphen and acronym |
| Search Terms for MOO studies | Multi-objective optimization; Tri-objective optimization or Bi-objective optimization with each of these with/without hyphen and optimisation (instead of optimization) |

Important details of the search adopted for finding studies on MCDM applications, are as follows.

**Database and Keywords**: We conducted a systematic search using the SCOPUS database, employing search words such as MCDM, MADM, MCDA or MCA with each of these as full form with/without hyphen and acronym (Table 5), in the title, abstract and/or keywords, to capture a comprehensive set of relevant articles. For MOO, we used keywords including Multi-objective optimization; Tri-objective optimization or Bi-objective optimization with each of these with/without hyphen and optimisation (instead of optimization).

**Timeframe**: We limited the search to articles published between 2000 and 2023 to reflect recent and relevant advancements.

**Subject Area and Document Type**: Searches were refined to focus on chemical engineering journals. Document type is limited to journal articles.

**Screening**: We screened the titles of journals in the chemical engineering subject area (as categorized by Scopus) and excluded those covering all engineering disciplines or science. Some examples of excluded journals are Applied Sciences Switzerland, Cogent Engineering, International Journal of Molecular Sciences, International Journal of Mechanical and Production Engineering Research and Development, Defence Science Journal, Process Biochemistry, Chemistry A European Journal, Journal of Biotechnology, and Journal of Science and Engineering. This screening is to ensure each article is related to MCDM applications in chemical and process engineering.

The SCOPUS search with the terms for MCDM studies (Table 5) identified 931 articles published in the period: 2000 to 2023. As shown in Figure 6, applications of MCDM in chemical engineering have been increasing steadily. In particular, the increase is significant from 2020, with more than 80 journal articles in a year. There are studies (including some of our papers) that employed one or more MCDM methods or investigated selection of one of the alternatives but did not contain the search terms used for MCDM studies, in the title, abstract or keywords. Hence, they are not part of 931 articles found by the search for MCDM studies. So, there are more relevant papers not captured by the present search. For comparison with a closely related area, another search was performed using the search terms for MOO studies in Table 5, in the same 160+ journals searched for MCDM studies. This search gave a total of 2753 journal articles, which is thrice the number for MCDM studies. This trend of more studies on MOO compared to those on MCDM can be seen throughout the search period (Figure 6).

Table 6 lists 23 journals in the chemical engineering area that published 10 or more articles on MCDM studies from the year 2000 to 2023. The total number of articles covered by these 23 journals is 477 or about 50% of 931 articles found by the MCDM search. The rest of 50% is from the remaining more than 140 journals that published 1 to 8 articles during the search period. Almost all journals in Table 6 are within the scope of chemical engineering. Specifically, the journals with high number of articles are Processes (with 64 articles), Journal of Environmental Planning and Management (with 41 articles) and Chemical Engineering Transactions (with 33 articles).

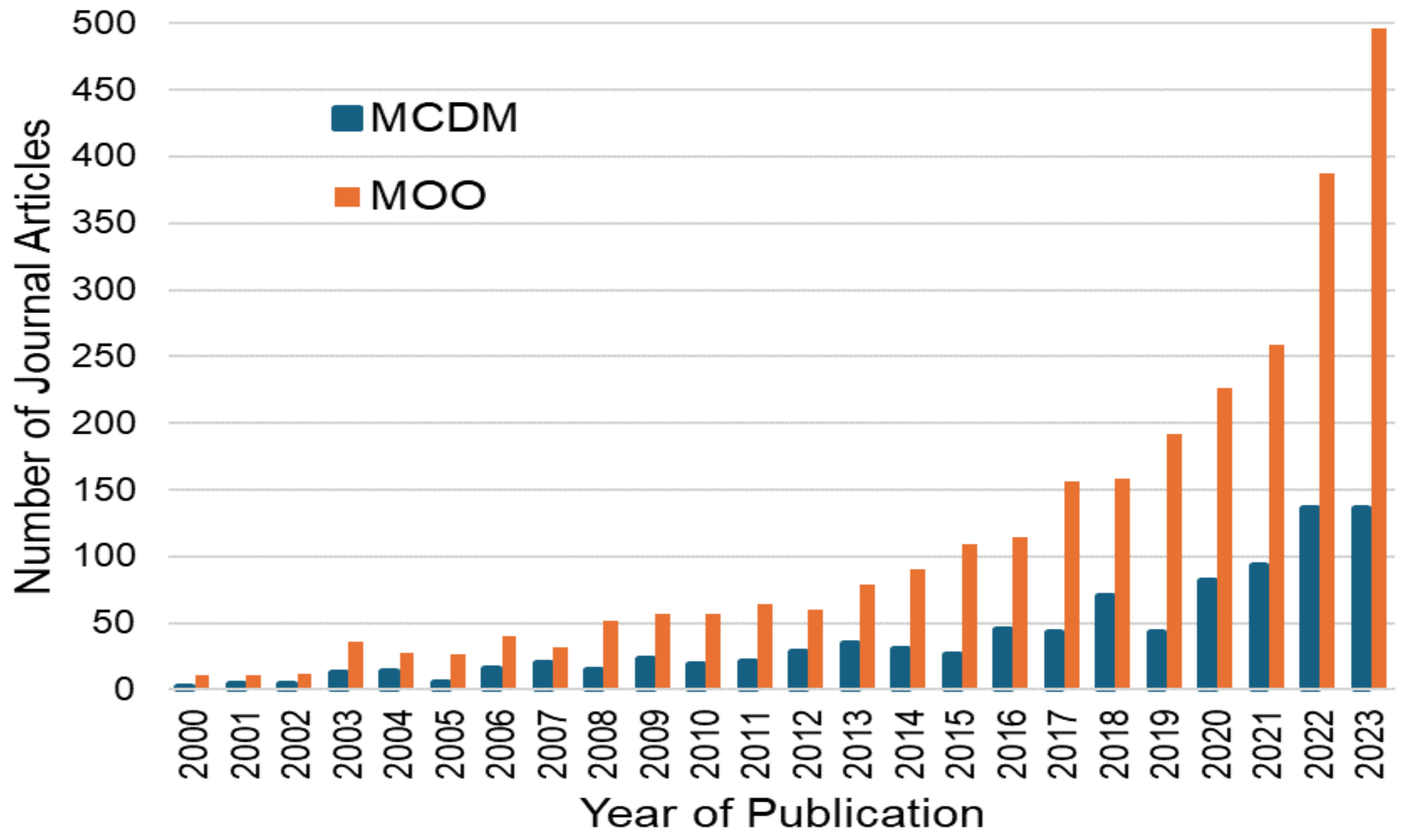

Figure 6. Number of journal articles on MCDM and MOO in chemical engineering, from the year 2000 to 2023

Table 6. Journals that published 10 or more articles on MCDM from the year 2000 to 2023

| No. | Journal Title | Number of Articles |
|---|---|---|
| 1 | Processes | 64 |
| 2 | Journal of Environmental Planning and Management | 41 |
| 3 | Chemical Engineering Transactions | 33 |
| 4 | Fuel | 26 |
| 5 | Process Integration and Optimization for Sustainability | 26 |
| 6 | Process Safety and Environmental Protection | 24 |
| 7 | Computers and Chemical Engineering | 22 |
| 8 | International Journal of Sustainable Energy | 22 |
| 9 | Desalination | 20 |
| 10 | Energy Sources Part B: Economics Planning and Policy | 20 |
| 11 | Journal of Loss Prevention in The Process Industries | 18 |
| 12 | Biofuels Bioproducts and Biorefining | 17 |
| 13 | Bioresource Technology | 17 |
| 14 | Chemical Engineering Journal | 17 |
| 15 | Fibers and Polymers | 16 |
| 16 | Industrial and Engineering Chemistry Research | 15 |
| 17 | ACS Sustainable Chemistry and Engineering | 14 |
| 18 | Chemometrics and Intelligent Laboratory Systems | 12 |
| 19 | Computer Aided Chemical Engineering | 12 |
| 20 | Environmental Progress and Sustainable Energy | 11 |
| 21 | Bioresources | 10 |
| 22 | Ceramics International | 10 |
| 23 | Journal Of Colloid and Interface Science | 10 |

Table 7 recognizes the researchers who published 4 or more articles during the search period of 2000 to 2023. The total number of researchers in this table is 27, and their publications total to 155 (out of 931 found by the search for MCDM studies). At least one paper from each of these active researchers is reviewed in the next section of this paper. In all, researchers who studied MCDM in chemical engineering come from more than 80 countries, which shows that MCDM studies in chemical engineering are widespread throughout the world. The top 5 countries with high number of contributors to MCDM studies in chemical engineering are China, India, Iran, United States America and Turkey with 181, 106, 76, 73 and 54 articles, respectively. In total, these articles account for 490 or about 53% of 931 found by the SCOPUS search for MCDM studies.

Table 7. Researchers, who published 4 or more journal articles on MCDM from the year 2000 to 2023

| Names of Researchers | Number of Articles |
|---|---|
| Boran, F.E.. Ren J. | 9 |
| Promentilla, M.A.B. | 8 |
| Kalita, K.; Stuart, P.R.; Wang, C.N. | 7 |
| Aviso, K.B.; Boran, K.; Dong, L.; Feizizadeh, B.; Hasanzadeh, R.; Kokot, S.; Linkov, I.; Majumdar, A.; Shen, W.; Tan, R.R. | 6 |
| Azdast, T.; Farid, S.S.; He, C.; Wang, Z. | 5 |
| Gun'ko, V.M.; Moghassem, A.R.; Polatidis, H.; Rangaiah, G.P.; Shanmugasundar, G.; Todeschini, R.; Yang, A. | 4 |

As described in Section 2 and depicted in Figure 2, MCDM procedure involves the steps of normalization, weighting and ranking, and there are different methods for each of these steps. Each MCDM method is associated and generally used with a particular normalization method, but this is not so with the weighting method (i.e., different weighting method can be selected for use with a MCDM method). To find chemical engineering studies using common weighting methods, search for each of the chosen weighting methods was conducted within the 931 journal articles found on MCDM studies. This type of search for the entropy, BWM, CRITIC, variance and standard deviation methods found 95, 39, 30, 22 and 12 respectively. These results show that the entropy method for weights is the most popular for MCDM in chemical engineering. It is followed by BWM and CRITIC methods.

Next, to learn about the use of MCDM methods in chemical engineering applications, a separate search was conducted for each of the methods that are thought to be popular. For this, settings for search fields, publication year, subject area, document type and language are the same as those in Table 5. However, search terms are different, and these are listed in Table 8 along with the number of journal articles retrieved by the search for each MCDM method. Note that these searches for articles on using MCDM methods are not within the 931 journal articles found on MCDM studies. The reason for this is to capture those articles used and mention a particular MCDM method in the title, abstract and keywords but do not include the search terms used for MCDM studies (stated in Table 5).

Note that articles that used a MCDM method but did not state that method in the title, abstract and keywords are not captured in the searches for studies using different MCDM methods. Hence, there may be more articles that employed an MCDM method in chemical engineering studies, in addition to the search results in Table 8. According to this table, the most popular MCDM method used in chemical engineering is AHP, which occurs in 738 articles. As outlined in the earlier sections, AHP can be used for criteria weights and also as a MCDM method. Hence, these 738 articles might have used AHP for criteria weights or ranking the alternatives. AHP is followed by TOPSIS and GRA with nearly equal numbers of articles (368 and 367). The fourth most popular MCDM method in chemical engineering is ANP (with 166 articles). These findings are not surprising since these four methods are some of the earliest methods, which were proposed about 40 years ago. The MCDM methods: PROMETHEE, VIKOR, SAW (or WSM), Goal Programming, CODAS and LINMAP have occurred in 55 to 29 articles in chemical engineering (Table 8). On the other hand, the number of journal articles found for each of ELCTRE, FUCA, MOORA, COPRAS, MEP/WPM and MABAC, is less than 20.

Table 8. MCDM methods and the number of journal articles in chemical engineering using them, during the period: 2000 to 2023

| No. | MCDM Method and Search Terms Used | Number of Journal Articles |
|---|---|---|
| 1 | Analytic Hierarchy Process or AHP | 738 |
| 2 | Technique for Order of Preference by Similarity to Ideal Solution or TOPSIS | 368 |
| 3 | Gray Relational Analysis or Grey Relational Analysis or GRA | 367 |
| 4 | Analytic Network Process or ANP | 166 |
| 5 | Preference Ranking Organization Method for Enrichment Evaluation or PROMETHEE | 55 |
| 6 | Viekriterijumsko Kompromisno Rangiranje or VIKOR | 46 |
| 7 | Simple Additive Weighting (SAW) or Weighted Sum Method (WSM) | 42 |
| 8 | Goal Programming | 41 |
| 9 | Combinative Distance-Based Assessment or CODAS | 30 |
| 10 | Linear Programming Technique for Multidimensional Analysis of Preference or LINMAP | 29 |
| 11 | Elimination and Choice Translating Priority or ELECTRE | 19 |
| 12 | Faire Un Choix Adéquat or FUCA | 14 |
| 13 | MOO on the Basis of Ratio Analysis or MOORA | 12 |
| 14 | Complex Proportional Assessment or COPRAS | 8 |
| 15 | Multiplicative Exponent Weighting (MEP) or Weighted Product Model or Weighted Product Method (WPM) | 6 |
| 16 | Multiattributive Border Approximation Area Comparison or MABAC | 3 |

As part of this research, bibliometric review was performed using VOSviewer software version 1.6.20. This free program can be used to create maps of authors, their countries, keywords etc. The input data file for this software is Scopus search data, which is exported in csv format from Scopus. Figure 7 shows the co-authorship map of different countries. The size of the circle of a country indicates its greater participation in research on MCDM applications in chemical engineering. The number of links (edges) of a country with other countries shows the level of cooperation, and the strength of each cooperation is shown by the thickness of the connecting edges (i.e., a thicker line shows a stronger link and vice versa). According to Figure 7, researchers from China (14.1% out of the articles found by the search), India (8.5%), Iran (6.6%), United State (6.1%), Turkey (4.3%), United Kingdom (3.7%) and Canada (3.6%) have contributed the most articles. Further, China, United state, India and Iran have more citations. In terms of the strength of cooperation, China, Saudi Arabia and Iran had more link strength (Figures 8a to 8c).

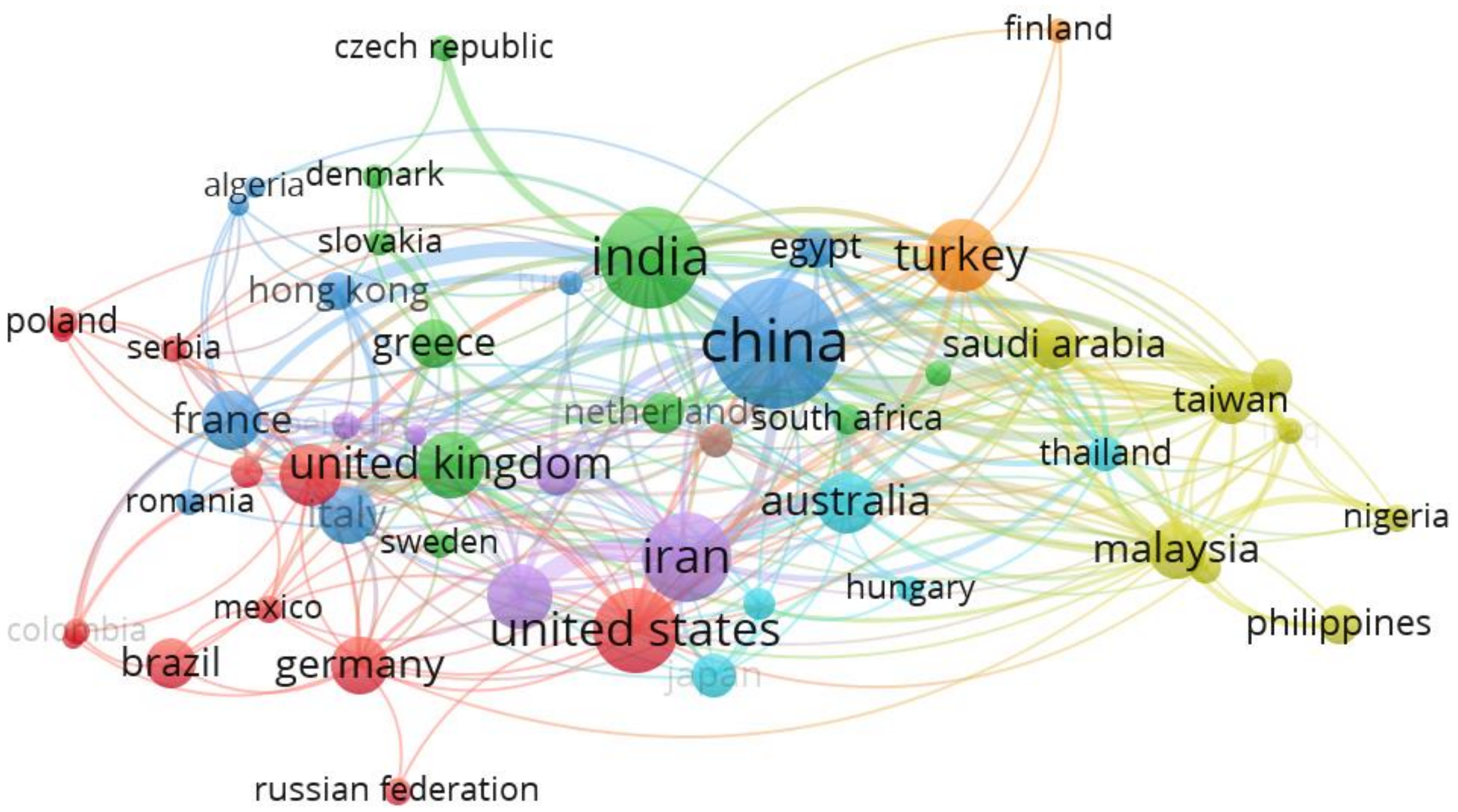

Figure 7. Bibliographic coupling of journal articles based on countries using VOSviewer network visualization. For clarity, countries that contributed 5 or fewer articles are excluded.

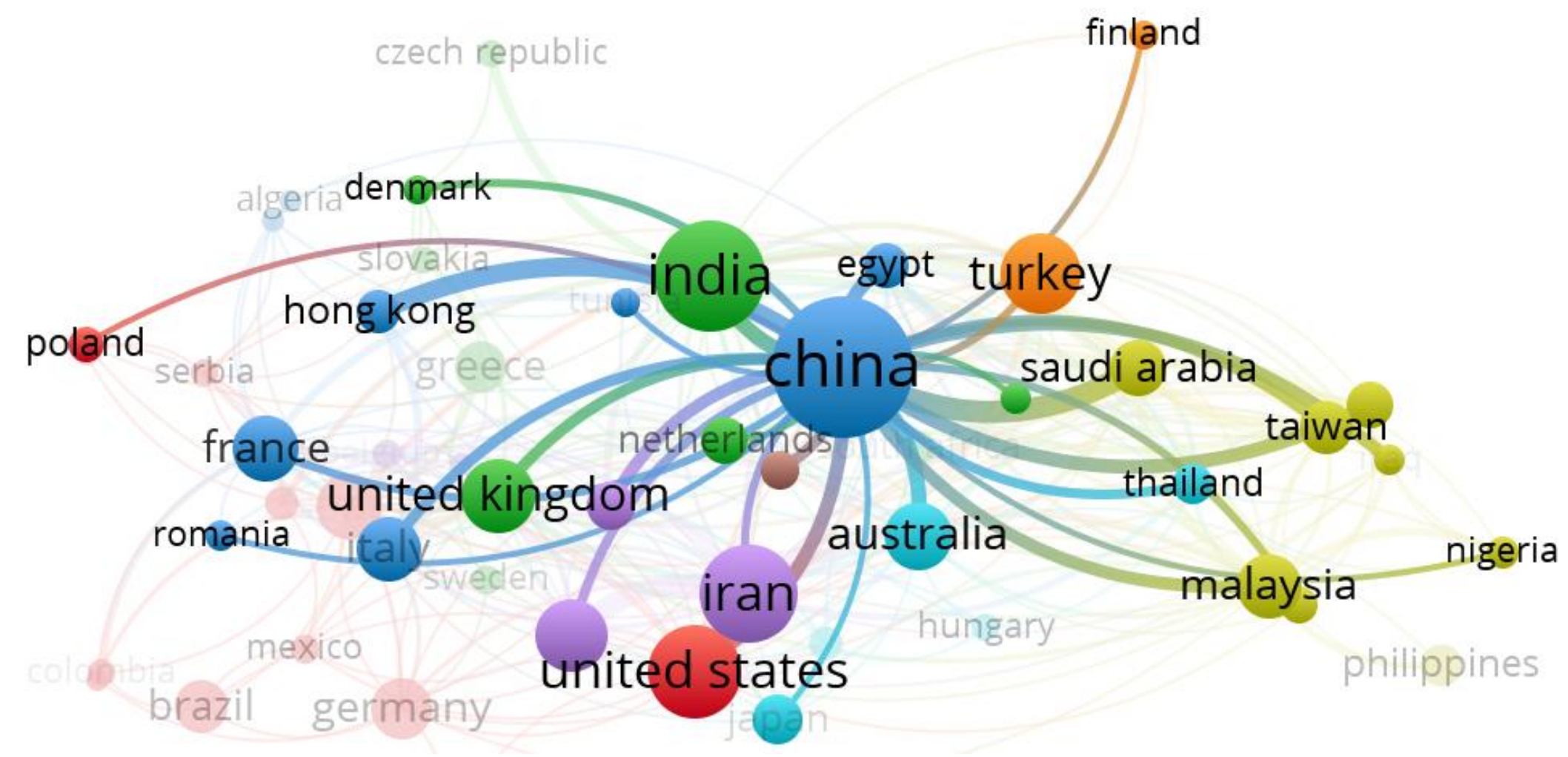

Figure 8a. Total link strength of China with other countries

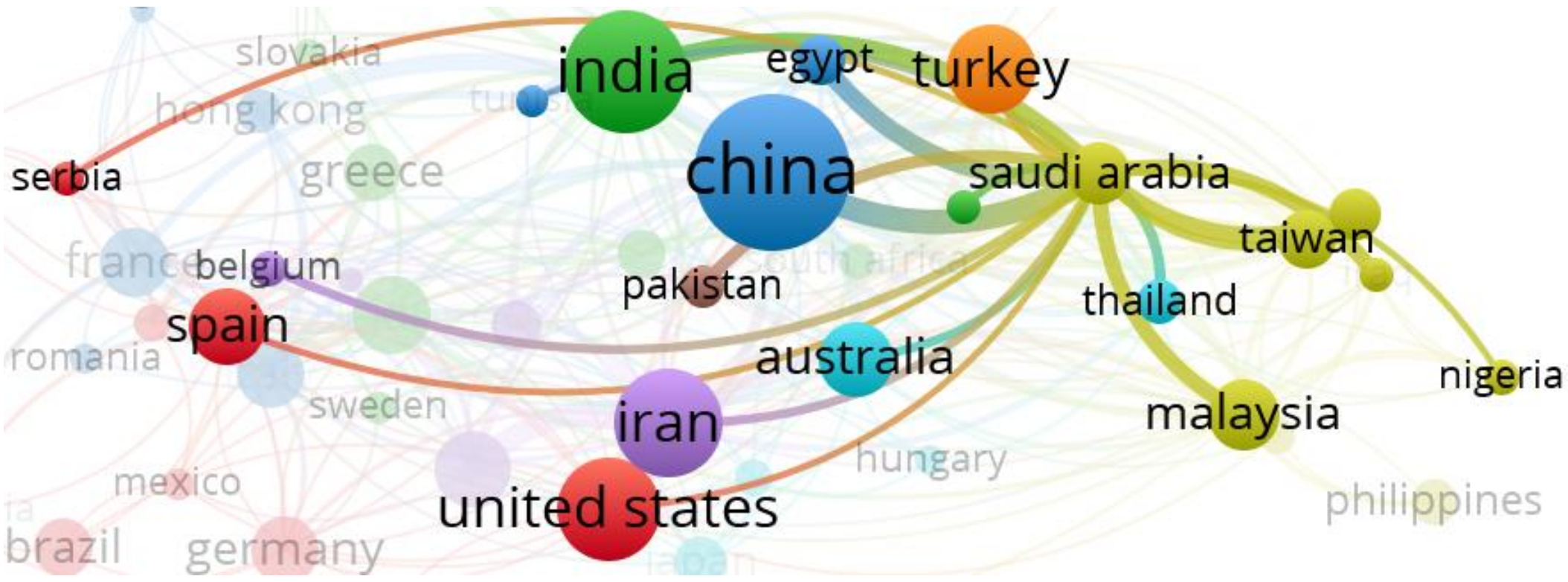

Figure 8b. Total link strength of Saudia Arabia with other countries

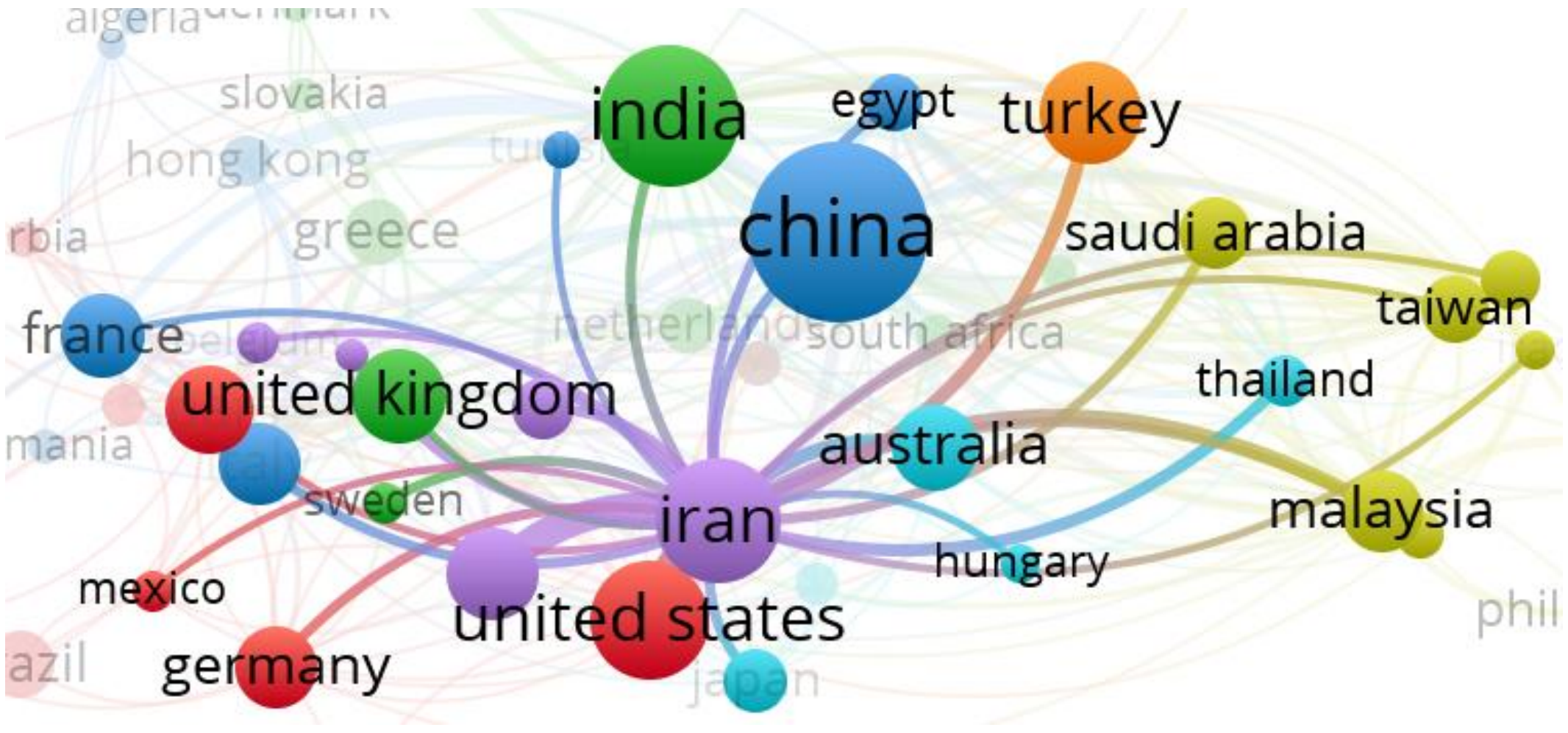

Figure 8c. Total link strength of Iran with other countries

The journal articles found on the application of MCDM methods in chemical engineering were further investigated by analyzing the relationship between keywords related to chemical engineering with those related to MCDM, utilizing the VOSviewer program. Figure 9 illustrates the co-occurrence of authors' keywords in the published work. For preparing this figure, minimum number of occurrences of a keyword was set at 5. As expected, "multi-criteria decision making" and closely related phrases are keywords in many papers. MCDM methods such as TOPSIS, entropy and fuzzy AHP can also be seen in Figure 9, which indicate their frequent occurrence (compared to others) in authors' keywords.

Figure 10 shows the degree of association of four keywords: "sustainability" (Figure 10a), "life cycle assessment" (Figure 10b), "renewable energies" (Figure 10c) and "desalination", (Figure 10d), with keywords related to MCDM. The number of links for the sustainability keyword is 23 and the total link strength (TLS) is 39. These are 18 links and 24 TLS for the life cycle assessment keyword, 17 links and 20 TLS for the renewable energy keyword, and 13 links and 16 TLS for the biorefineries keyword. These numbers indicate the greater application of MCDM methods in these areas of chemical and process engineering.

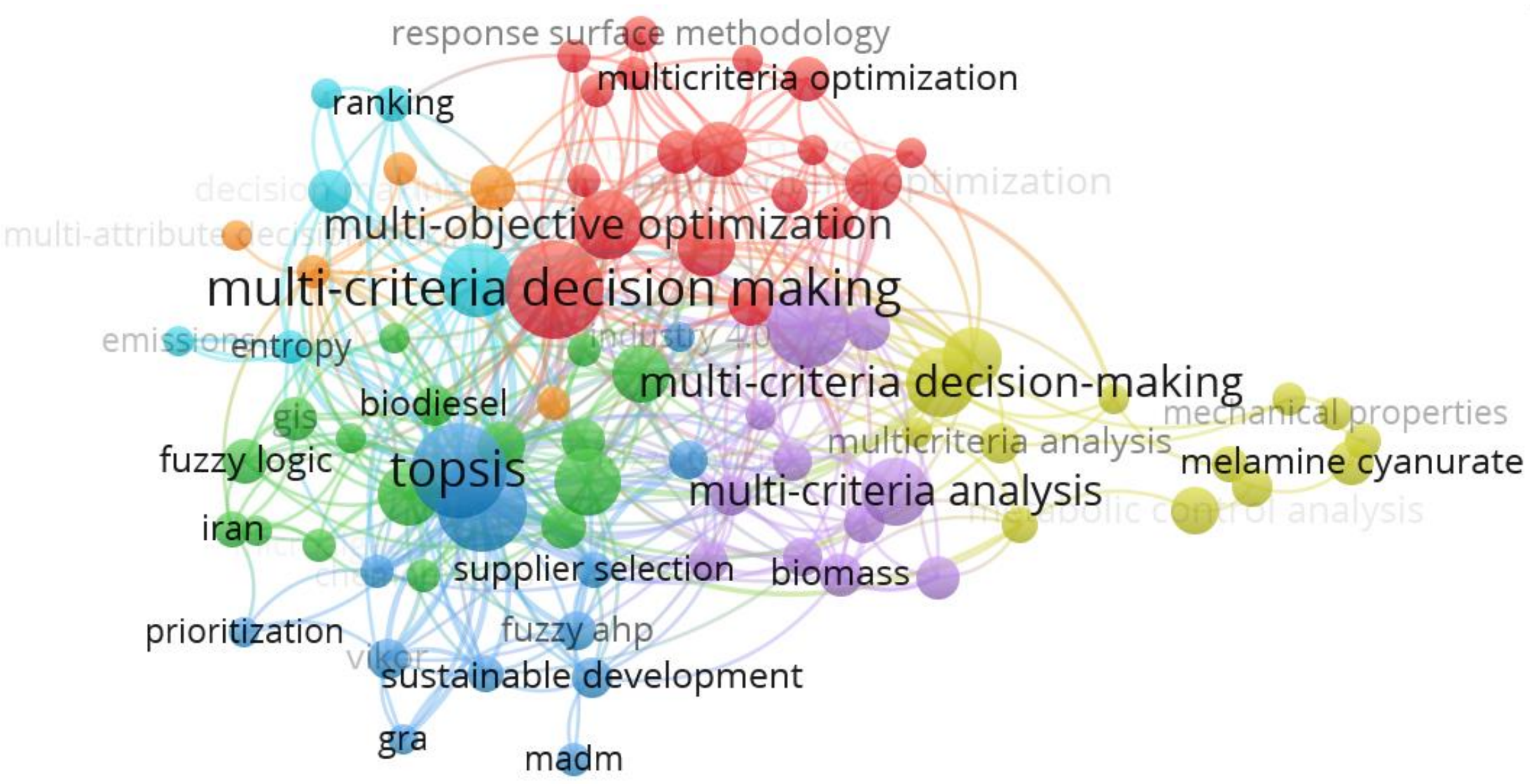

Figure 9. Co-occurrence of authors' keywords in the articles found on MCDM applications in chemical engineering, using VOSviewer

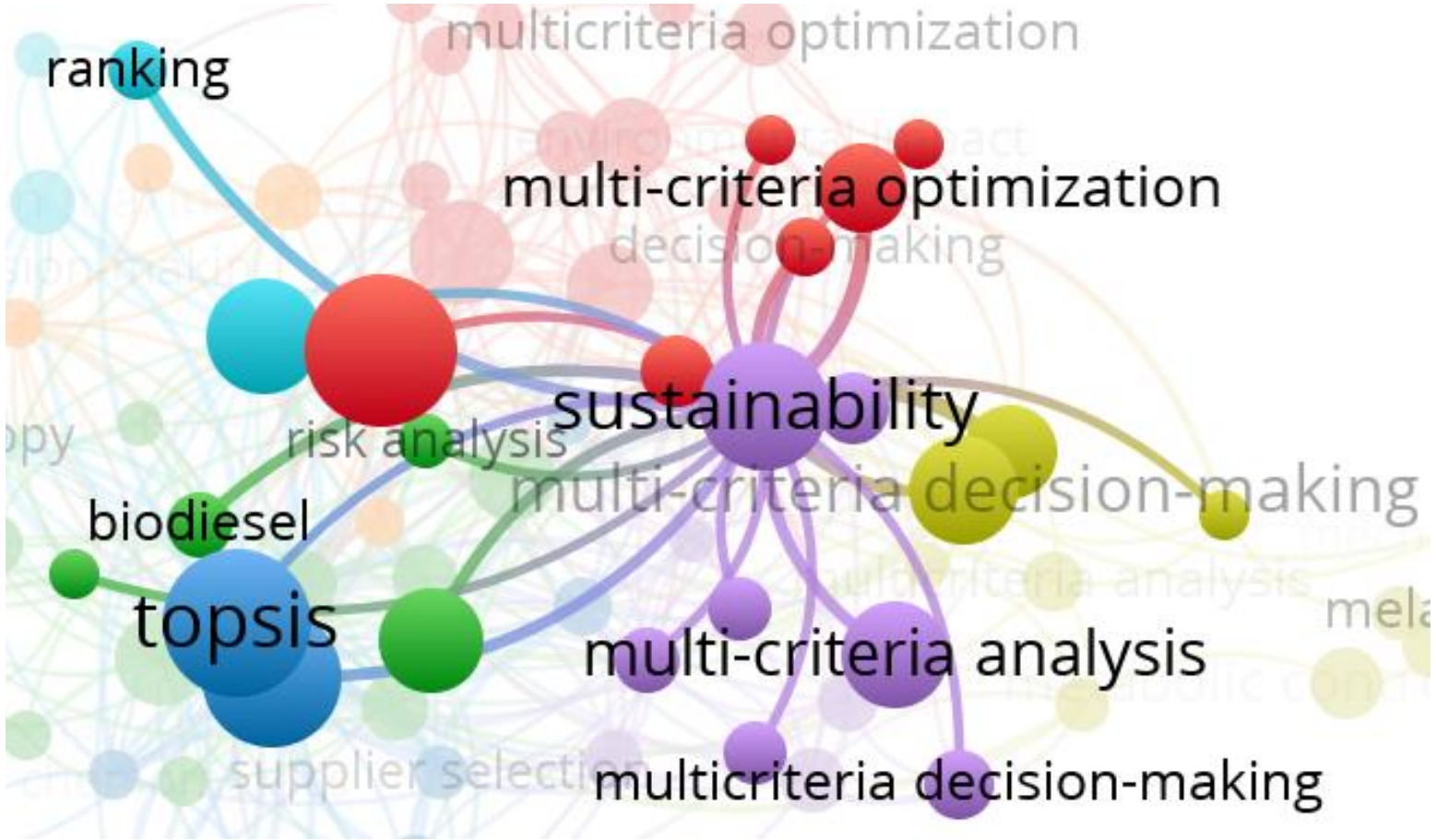

Figure 10a. Link of "sustainability" with MCDM keywords

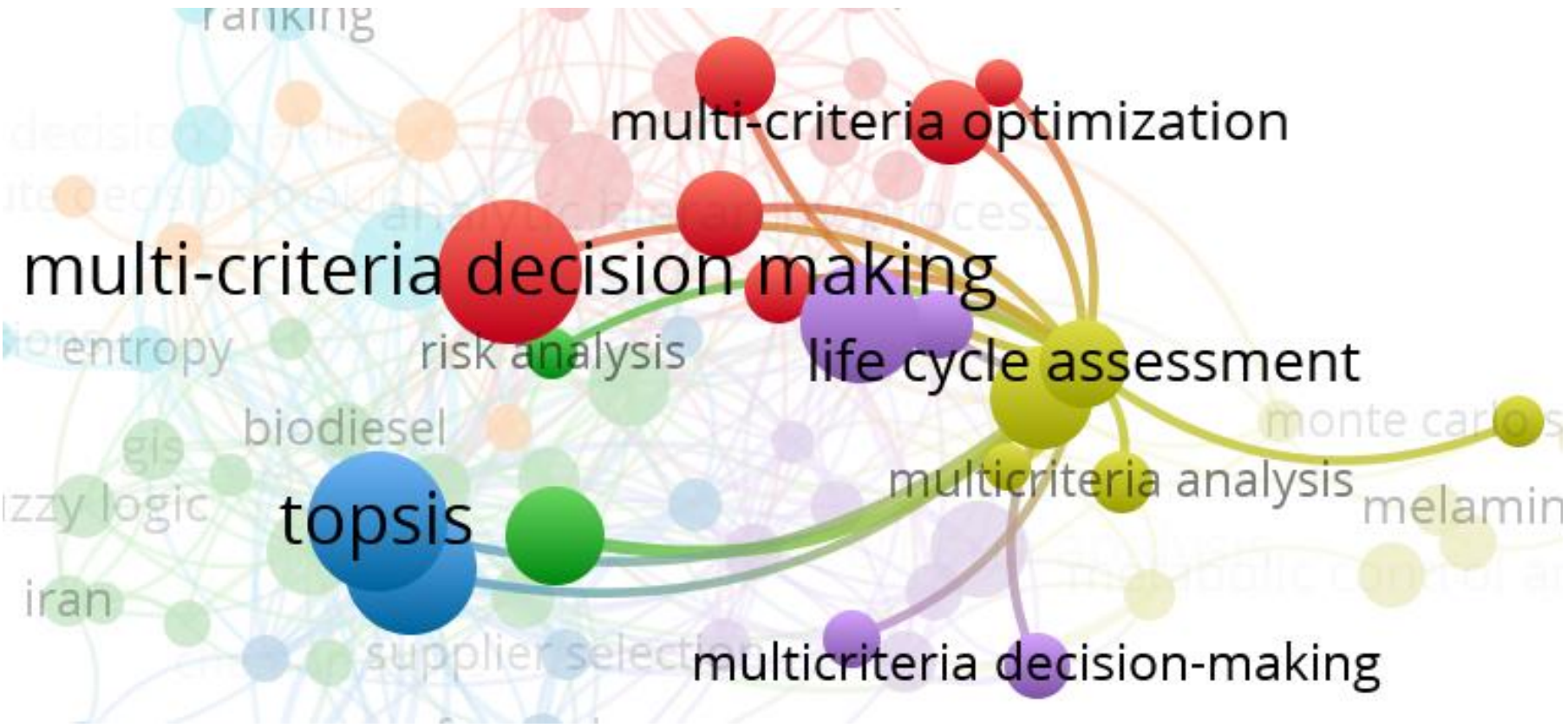

Figure 10b. Link of "Life cycle assessment" with MCDM keywords

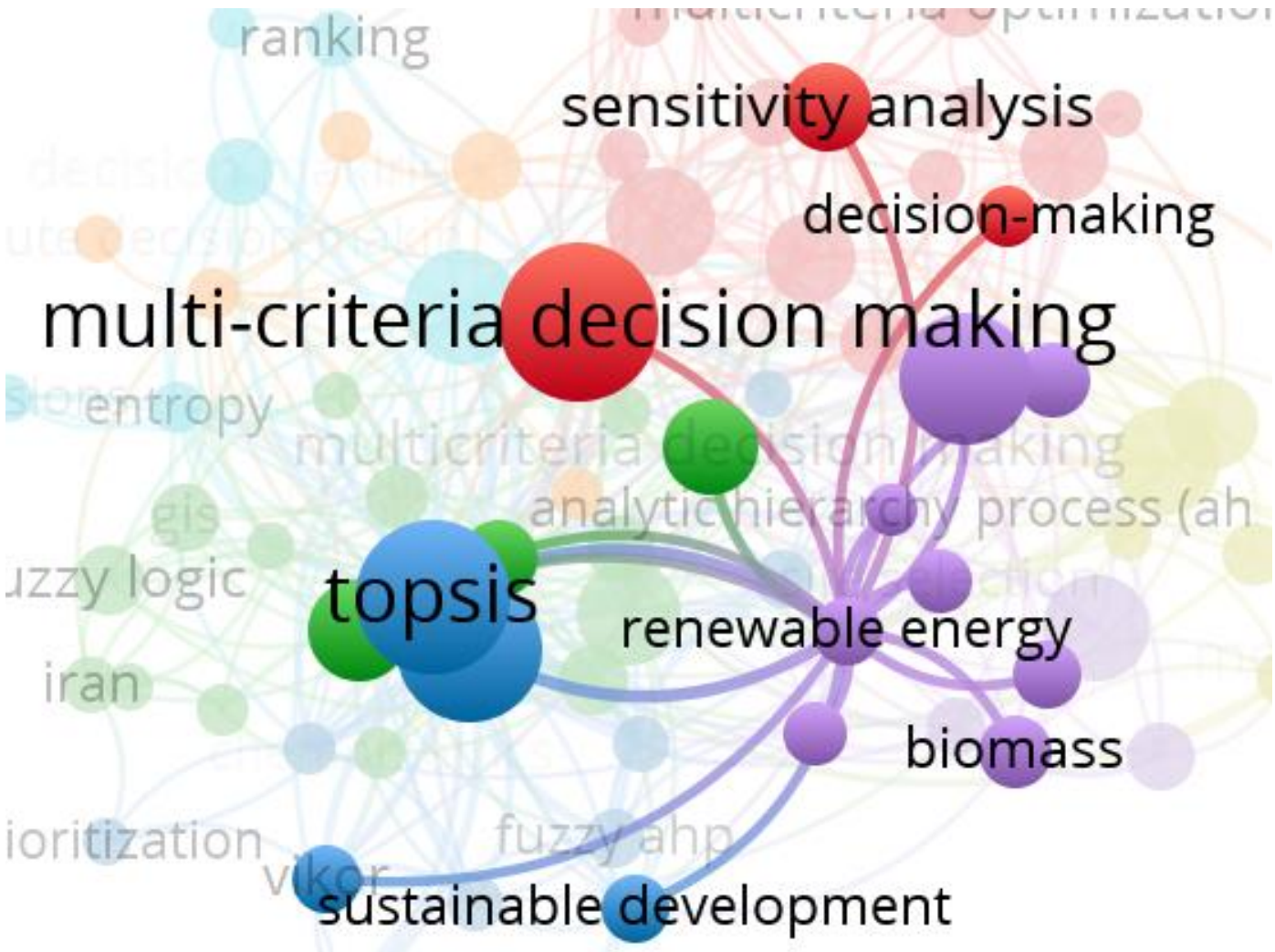

Figure 10c. Link of "renewable energy" with other related MCDM keywords

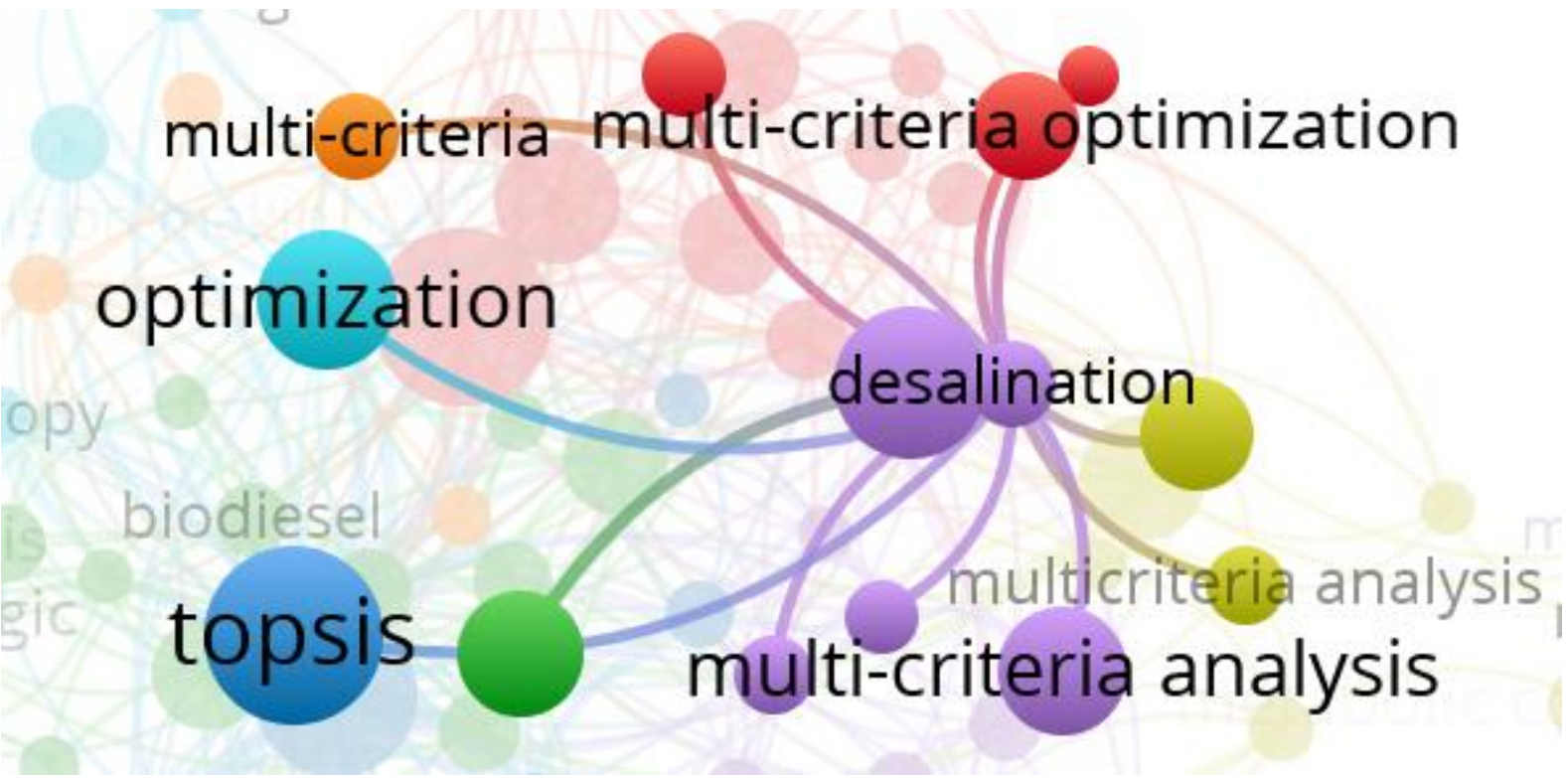

Figure 10d. Link of "desalination" with other related MCDM keywords

We acknowledge a few limitations in our approach to literature review and selection. First, our review is limited to journal articles indexed in the SCOPUS database, which, while extensive, may omit relevant studies indexed in other databases. Second, our focus is specifically on the chemical and process engineering domain; for this, we chose the subject area of Chemical Engineering in Scopus and manually excluded some journals covering all engineering disciplines or science (including chemistry). While this scope and focus are sufficient to cover more than 160 journals and for a comprehensive review of MCDM applications in chemical engineering, they may not cover all journals in the related areas such as bioprocess engineering, energy, food processing, membrane technology and pharmaceutical engineering.

**8. Overview of Selected Journal Articles on MCDM Studies**

Selected journal articles on MCDM studies in chemical engineering are summarized in chronological order, in Table 9. As stated earlier, these articles cover at least one paper by the active researchers with 4 or more articles on MCDM listed in Table 7. These articles and 5 highly cited articles are briefly reviewed in the following paragraphs.

Table 9. Summary of selected journal articles on MCDM studies in chemical engineering

| Reference | Application | No. of Criteria | No. of Alternatives | Weighting Method(s) | MCDM Method(s) | Program/ Software |
|---|---|---|---|---|---|---|
| Ni et al. (2011) | Ranking the near-infrared spectrum in the analysis of potato crisps | 4 | 84 | Not stated | PROMETHEE and GAIA | Decision Lab 2000 software |
| Valipour and Moghassem (2014) | Selecting suitable yarn for Weft knitting process | 7 | 48 | Manual weights | PROMETHEE | Not stated |
| Majumdar and Singh (2014) | Determination of the quality value of cotton fibers | 5 | 17 | Optimal weight found by GA | TOPSIS | Not stated |
| Grisoni et al. (2015) | Eight datasets of different type from the literature | 3 to 27 | 7 to 86 | Different weighting methods | Modified Hasse diagram | Not stated |
| Polatidis et al. (2015) | Comparison of ELECTRE III and PROMETHEE II for a geothermal application | 5 | 4 | Manual weights | ELECTRE III and PROMETHEE II | Not stated |
| Boran (2017) | Evaluation of power plants in Turkey | 5 | 5 | Manual weights | Fuzzy TOPSIS | Not stated |
| Yang et al. (2017) | Formulation conditions for high-concentration monoclonal antibodies | 2 | 9 | Manual weights | SAW (WSM) | Not stated |
| Boran (2018) | Evaluation of renewable energy resources | 12 | 5 | AHP | Fuzzy VIKOR | Not stated |
| Jenkins & Farid (2018) | Cost-effective bioprocess design for the manufacture of allogeneic CAR-T cell therapies | 7 | 7 | Various weights | Additive weighting technique (like SAW) | Coding in Visual Basic for Application (VBA) |
| Ren et al. (2018) | Life cycle sustainability assessment under uncertainties | 14 | 5 | Interval BWM | Interval TODIM | Not stated |
| Rycroft et al. (2018) | Generic example | 4 | 4 | Not stated | Not stated | Not stated |
| Ongpeng et al. (2020) | MCA in energy retrofit of buildings | 5 | 3 | AHP | VIKOR | Not stated |
| Bernardo et al. (2022) | Hydrogen solid-state storage | 4 | 4 | Manual weights | Fuzzy MCDM | Not stated |
| Hasanzadeh et al. (2022) | Waste polymeric foam gasification | 3 | 2 | Manual weights | TOPSIS | Not stated |
| Liu et al. (2022) | Sludge valorization process for value-added products | 11 | 4 | Fuzzy BWM | Fuzzy PROMETHEE II | Not stated |
| Wang et al. (2022a) | Sustainable supplier selection with technology 4.0 evaluation | 12 | 5 | Ordinal priority approach (OPA) | MARCOS | Not stated |
| Bele et al. (2023) | Production of renewable and sustainable biofuels | 4 | 2 | Not stated | Not stated | I-BIOREF software developed by Natural Resources Canada |
| Feizizadeh et al. (2023) | Forest fire risk modeling | 13 | Not applicable | ANP | Not applicable | Not applicable |
| Hasanzadeh et al. (2023) | Plastic waste gasification | 5 | 7 | Manual weights | TOPSIS, GRA | Not stated |
| Shanmugasundar et al. (2023) | Selection of Industrial Robots | Case I (3) Case II (5) Case III (6) | Case I (5) Case II (7) Case III (4) | Mean, Standard deviation, Entropy, Preference selection index (PSI) methods | SAW, TOPSIS, LINMAP, VICOR, ELECTRE-III, NFM | Not stated |
| Sun et al. (2023) | Design of side-stream extractive distillation for separation of azeotropic mixtures | 3 | 21 obtained by MOSPSO | Entropy method | TOPSIS | Coding and calculations in MATLAB |
| Wang et al. (2023c) | Study of MPC using ML, MOO and MCDM | 3 | 30 Found by NSGA-II | CRITIC | PROBID | Coding and calculations in Python |
| Yang et al. (2023) | Design of extractive dividing wall column | 3 | 30 obtained by MOPSO | Not needed | GRA | Coding and calculations in MATLAB |

Ni et al. (2011) analyzed NIR spectra of potato chips (i.e., a complex product) with the help of PROMETHEE and GAIA (Geometrical Analysis for Interactive Aid). Four similar types of the 'original flavor' potato chips from different manufacturers were chosen, and there were 20 samples for each type. Apart from the collection of the NIR data, the samples were analyzed for 4 quality parameters, i.e., fat, moisture, acid and peroxide values of the extracted oil by standard methods. The data matrix consisted of 84 potato chips as the alternatives and 4 mean NIR spectral objects of the groups (Aavg, Bavg, Cavg and Davg) as the criteria. The alternatives were compared according to the PROMETHEE outflow φ rank order. Results show that, quantitatively and qualitatively, the spectral objects tended to separate into the four groups of potato chips but there was some overlapping of samples.

Valipour and Moghassem (2014) also applied PROMETHEE to select suitable drawing frame variables for 30 Ne rotor spun yarn intended to be used in Weft knitting process. The performances of three variables in a drawing frame were evaluated based on 7 quality parameters (criteria) of 48 rotor spun yarns, using PROMETHEE. Sensitivity analysis was also performed to assess the stability of the final ranking.

Majumdar and Singh (2014), proposed a new algorithm based on the combination of TOPSIS and genetic algorithm (GA). The proposed method was applied to determine the quality value of cotton fiber considering two yarn properties, namely, yarn tenacity and unevenness. The weights of the cotton fiber properties were optimized by GA whereas TOPSIS helps to select cotton fiber with the best quality value.

Grisoni et al. (2015) proposed a modified version of Hasse diagram, which is useful for partial order ranking of importance for MCDM, to reduce the number of incomparabilities and derive weighted rankings of alternatives (termed as objects in the paper). They demonstrated the effectiveness of the modified version on 8 datasets (of which 4 are of chemical and environmental interest, 2 on bibliometric/journal metrics and 2 on comparison of classification methods and cities).

Polatidis et al., (2015) compared ELECTRE III and PROMETHEE II methods using alternative investment scenarios for a particular geothermal field. Five criteria, namely, TOE (tons of oil equivalent) saved/yr, environmental impact, jobs created, return on investment and risk were employed, and four scenarios were considered. The data for these were taken from the authors' earlier study. Both ELECTRE III and PROMETHEE II methods gave similar results on ranking the four scenarios in this case study.

Boran (2017) used fuzzy TOPSIS method to evaluate the power plants in Turkey. Fuzzy TOPSIS was applied due to incomplete information in the evaluation process. Five power plants (fossil fuels-based, geothermal, wind, hydro, and gas-based power plants) were selected as alternative power plants. Five criteria: efficiency (%), installation cost (USD/kW), electricity cost (USD/kWh), emission of carbon dioxide (kg$CO_2$/kWh), and social acceptance were considered in the evaluation process. Hydro-based power plants are the top choice due to their low emissions of CO2.

Yang et al. (2017) employed MCDM to identify the optimal formulation conditions for producing high-concentration monoclonal antibodies. In one framework, they used WSM (SAW) for two criteria, namely, viscosity and aggregation scores to rank 9 formulation conditions, and also analyzed the effect of weights to find the optimal formulation conditions.

Boran (2018) evaluated the renewable energy resources, namely, wind, hydro, solar, geothermal and biomass, for Turkey using intuitionistic fuzzy VIKOR method. For the evaluation, 4 main criteria – technological, environmental, sociological and economic, and 12 sub-criteria were used; these involved both quantitative and qualitative criteria. The sensitivity analysis shows that wind power is the first-ranked option in 14 out of 24 scenarios and the second-ranked option in 9 out of 24 scenarios.

Jenkins and Farid (2018) developed a decision tool to evaluate 7 bioprocess flowsheets for the manufacture of an allogeneic CAR-T cell therapy, from an economic and operational perspective. The flowsheets contain different bioprocess technologies, namely, T-flasks, gas permeable vessels, rocking motion bioreactors, an integrated processing platform, magnetic-activated cell sorting platform (MACS) purification and spinning membrane filtration technology. They were evaluated with 7 financial and operational criteria: cost of goods per dose, fixed capital investment, ease of operation, process control, validation effort, ease of scale-up and process containment. MCDM with sensitivity analysis showed that the flowsheet consisting of rocking motion bioreactor, spinning filter membrane, and a standalone MACS platform as the preferred process design for the criteria considered.

Ren et al. (2018) developed a novel MCDM method (namely, interval TODIM, an acronym in Portuguese for interactive and multicriteria decision making), which can address the decision-making matrix with interval numbers, to prioritize industrial systems under data uncertainties. The interval BWM was developed by extending the traditional BWM to interval conditions. Five alternative industrial process routes (namely, pulverized coal, combined cycle gas turbines, nuclear-pressurized water reactor, offshore wind powder-based electricity and solar-photovoltaics) for electricity generation in UK were studied. A total of 14 criteria including 4 economic criteria (capital cost, operation and maintenance cost, fuel cost, and total cost), 6 environmental criteria (global warming potential, ozone depletion, acidification potential, eutrophication, photochemical smog, and land occupation), and 4 societal criteria (social acceptability, employment, human toxicity potential, and total health impact from radiation) were employed for life cycle sustainability assessment of the chosen 5 routes for electricity generation. The offshore wind powder-based electricity was found to be the most sustainable; it is followed by combined cycle gas turbines, nuclear-pressurized water reactor, solar-photovoltaics and pulverized-coal in the descending order.

Rycroft et al. (2018) reviewed developments in nanomedical engineering from the year 2007 to 2016. Further, they outlined the role of MCDM to promote safety-by-design principles for developing nanomedicine. For this, a hypothetical case of 4 alternatives and 4 criteria covering safety, efficacy, cost of research & development and sustainability was used.

Ongpeng et al. (2020) evaluated energy retrofit strategies in buildings using environment, economic and technical performance criteria. Human health, ecosystem quality and resources are the environmental criteria whereas investment cost and energy potential (energy saved and generated from energy retrofit strategies) are economic and technical criteria, respectively. MCDM was carried out through AHP and VIKOR. The feasibility of net-zero energy buildings was achieved on an existing university building. However, MCDM showed that stakeholders can give more importance to the investment cost compared to the technical performance of retrofit strategies.

Bernardo et al. (2022) evaluated 4 nanoporous materials, namely, metal organic frameworks (MOFs), carbonaceous materials, metal hydrides, and complex hydrides using the fuzzy TOPSIS for hydrogen storage. Surface area, capacity, dehydrogenation temperature, and stability for hydrogen storage are the 4 selected criteria. The MOFs are the best alternative due to their relatively high surface area and excellent dehydrogenation temperature, whereas metal hydrides are the worst due to their relatively low surface area and sorption capacity.

Hasanzadeh et al. (2022) studied gasification of waste polyurethane foam. The processes were evaluated for the criteria: $CO_2$ emission, hydrogen and energy efficiencies. Effects of gasification temperature and moisture content in both air and steam gasification, and equivalence ratio in air gasification and steam to waste foam ratio in steam gasification were studied. The equivalence ratio is the actual fuel/air ratio to the stoichiometric fuel/air ratio. MCDM with different weights was investigated. The authors concluded that gasification with air has better performance compared to steam gasification in different scenarios.

Liu et al. (2022) analyzed 4 sludge valorization technologies: sludge anaerobic digestion with power generation and heat recovery, sludge incineration with electricity generation, sludge gasification for syngas and steam production, and supercritical water gasification for sludge treatment and syngas. These 4 processes were simulated in Aspen plus and then their sustainability assessment was conducted. Liu et al. used 11 criteria, which are of four types: environmental (i.e., climate change, acidification potential, human toxicity and eutrophication); economic (i.e., total capital cost, total operating cost and production sales), technological (i.e., energy efficiency, technical maturity and technology accessibility), and social (i.e., social acceptance). The MCDM showed that the sludge gasification for syngas and steam production is superior to the other 3 alternatives studied.

Wang et al. (2022a) applied ordinal priority approach (OPA) weighting method and fuzzy MARCOS to the selection of a sustainable supplier. Industry 4.0 technologies enable sustainable manufacturing, and thus sustainable supply chain management by reducing industrial waste and contaminants. They applied the procedure on five manufacturers of Vietnam's leather and footwear industry. The main criteria (with sub-criteria in brackets) were: logistics management (autonomous vehicles and robots, service level, blockchain, real-time manufacturing analytics system, smart containerization); production and operations management (3D printing, cloud computing, AI and ML, Internet of Things); and environmental competency (green product innovation, use of environmentally friendly technology, green image). The results show that green image, green product innovation, cloud computing, service level and blockchain are the significant criteria in evaluating sustainable practices in supply chains from the Industry 4.0 perspective.

Bele et al. (2023) compared 2 pathways: gasification combined with Fischer–Tropsch process and hydrothermal liquefaction (HTL) to produce biofuels, considering internal rate of return, capital investment, carbon intensity of biorefinery products, and total number of jobs created as the criteria. Of the two pathways, HTL was found to be attractive because of its outperformance from technical and environmental perspectives.

The focus of Feizizadeh et al. (2023) is on the application of artificial intelligence and geoinformation techniques to modeling forest fire risk. For this, they chose 13 criteria with 3 topographic-related, 3 related to land surface, 5 climate properties and 2 anthrophonic-related. Feizizadeh et al. determined weights for these criteria by ANP method, and then used them as independent variables in the multi-layer perceptron model for forest fire risk.

Hasanzadeh et al. (2023) used two agents (air and steam) for gasification of plastic waste and ranked 2 different plastic waste feeds (high- and low-density polyethylene, PE) using 5 criteria and two MCDM methods: TOPSIS and GRA. The criteria are: molar ratio of hydrogen to carbon monoxide in syngas, lower heating value of syngas, cold gas efficiency, exergy efficiency, and normalized carbon dioxide emission. Sensitivity analysis of ranking with respect to weights was carried out. Both TOPSIS and GRA showed that low-density PE feed has the best performance in the air gasification process. However, for steam gasification, TOPSIS and GRA ranked high-density PE and low-density PE as the best candidates, respectively.

Shanmugasundar et al. (2023) employed several MCDM and weighting methods (entropy, MW, SDV and PSI), for selecting an optimal robot depending on the application. Three different industrial robot selection problems (three cases) were studied using SAW, TOPSIS, LINMAP, VIKOR, ELECTRE-III and NFM. The entropy weighting method appears to be unique among the methods considered because of its tendency to be skewed in favor of or against certain criteria. The authors concluded that the MCDM methods seem to greatly rely on the weight allocation strategy, and so it is crucial for weights to properly reflect the relative importance of individual criteria without biases.

Sun et al (2023) used MOO and MCDM to design and intensify side-stream extractive distillation (SSED) for complex ternary azeotropic mixtures. Based on the existing triple-column extractive distillation (TCED) for a ternary mixture, three improved SSED processes including two SSED with one liquid side-stream (SSED-1 and SSED-2) and one SSED with double liquid side-stream (i.e., SSED-3) are proposed. Pareto-optimal solutions for the criteria: TAC, CO2 emission and process route index (PRI), were obtained using MOPSO, and then ranked using entropy weights and TOPSIS. Sun et al. concluded that introduction of the side-stream for extractive distillation could decrease exergy loss and increase thermodynamic efficiency. The top ranked SSED is recommended as the best alternative considering economy, environment and safety simultaneously.

Wang et al. (2023c) proposed a comprehensive ML aided model predicted control (MPC) with MOO and MCA methodology for chemical process control. NSGA-II was used to generate Pareto-optimal solutions (alternatives); it was followed by CRITIC weighting method and PROBID to select one of the optimal solutions. The proposed methodology was illustrated on a continues stirred tank reactor. It achieved the intended optimization for multiple objectives in MPC without compromising the closed-loop stability of the controlled system.

Yang et al. (2023), developed 2 intensified energy-efficient extractive distillation configurations for separating ethyl acetate and methanol from wastewater. The first one was an extractive dividing wall column with a side reboiler (EDWC-SR), which combines four columns into two. The second one was a double EDWC-SR (DEDWC-SR), which integrates four columns into a single unit. Simulation and MOO of the two configurations were performed using Aspen plus and MOPSO for economic (total annual cost), environmental ($CO_2$ emission), and safety (process route index) criteria. The obtained Pareto front having 30 alternatives was analyzed by GRA. Both EDWC-SR and DEDWC-SR have better economic and environmental performances relative to the base case but with a slightly lowered safety performance.

Citations (until the end of the year 2023) of each of the 931 articles found by the Scopus search vary significantly in the range of 0 to 270. This wide range is due to many factors such as research activity in the area, nature of the article (e.g., review or regular paper), publication year, journal impact, and weighting/MCDM methods employed. The 5 articles with more than 200 citations are briefly summarized below.

The most highly cited article is Moktadir et al. (2018) with 270 citations, published in the Process Safety and Environmental Protection journal. It investigates the challenges faced by the leather industry in Bangladesh in implementing Industry 4.0. Moktadir et al. identified 10 challenges (equivalent to criteria in MCDM) for ensuring environmental protection and control, and then determined their weights by BWM. The challenge, namely, the lack of technological infrastructure has the largest weight of 0.2284; so, it is concluded to be the most crucial challenge.

Feizizadeh and Blaschke (2013) studied the optimal utilization of land for dry-farm and irrigated agriculture in a region of Iran. They considered suitability factors (i.e., criteria in MCDM) related to soil, climatic conditions and water availability. AHP was used to determine weights for preparing suitability map layers. Research findings of Feizizadeh and Blaschke (2013) were provided for land use planning by the authorities.

The article of Carmody et al. (2007) is on the application of MCDM Methods: PROMETHEE and GAIA for performance evaluation of synthesized organo-clays and reference sorbents for adsorption of hydrocarbons, which is important for oil spill remediation. This study attracted 225 citations. Carmody et al. considered 13 sorbents of which 5 are organo-clay sorbents, and 9 performance criteria. They examined the effect of adding a criterion one by one on the ranking of 13 sorbents, initially based on 3 criteria related to hydrocarbon sorption capacity.

Pollock et al. (2013) investigated the potential of batch and continuous cell culture technologies to produce monoclonal antibodies. This article has been cited 215 times. These researchers considered 3 bioprocess alternatives and 9 performance criteria, which are grouped into 3 categories: operational, economic and environmental feasibility. MCDM of this application was performed by weighted sum method (like SAW).

Polatidis et al. (2006), which has 202 citations, developed a framework for selecting a MCDM method for renewable energy planning (REP) applications. This framework is comprehensive and considers the main characteristics of REP (such as environmental benefits, local impact, spatial and temporal distribution of costs and benefits, and many diverse stakeholders), main features of MCDM methods (such as ELECTRE, PROMETHEE, AHP and SAW) and prerequisites of MCDM methods for application to REP.

### 8.1. Discussion of Trends in the Articles Reviewed

Among the articles reviewed (Table 9), application of MCDM methods in some studies is to the Pareto-optimal solutions (e.g., Figure 1) found by solving a suitable MOO problem; in other words, ACM is based on these Pareto-optimal solutions. In many other applications, alternatives and criteria for ACM seem to be compiled from literature, survey of the market or experts (e.g., Table 1). In the 23 articles reviewed (Table 9), application of MCDM methods in chemical engineering is mostly in the material selection and process design areas. Various weighting methods have been employed in these papers, but manual weighting and AHP have been used more. Among the MCDM methods, TOPSIS (in 6 papers) and PROMETHEE II (in 4 papers), are popular in the 23 articles reviewed.

Almost all the studies in Table 9 employed only one weighting method and only one MCDM method. The few exceptions are: two or more weighting methos in Grisoni et al. (2015), Jenkins & Farid (2018) and Shanmugasundar et al. (2023); and more than one MCDM method by Ni et al. (2011), Polatidis et al. (2015) and Shanmugasundar et al. (2023). Specifically, Shamugasundar et al. applied 6 methods, namely, SAW, TOPSIS, LINMAP, VICOR, ELECTRE III and NFM for the selection of industrial robots. However, in each of these papers, there was only one application and so their findings may not be generalizable.

As summarized in Table 9, MCDM applications reported by researchers and practitioners in chemical engineering reflect the diversity of weighting and MCDM methods chosen and employed. Given this variety, it is quite challenging to say or propose that only certain MCDM methods should be applied in chemical and process engineering. Having said that, our previous works (Wang & Rangaiah, 2017; Wang et al., 2020; Nabavi et al., 2023) recommend CRITIC and entropy weighting methods, and several MCDM methods (TOPSIS, GRA, SAW, CODAS, and MABAC) based on simplicity of principle and algorithm, user inputs required and sensitivity to changes in ACM.

Finally, due to the ready availability of MCDM methods in open-source programs such as the MS Excel program and Python code (Section 6), we expect the increased use of MCDM in chemical and process engineering in the coming years. Further, we expect researchers to choose and employ several weighting and/or MCDM methods, as well as perform sensitivity analysis towards weighting/MCDM methods and possible changes in ACM (e.g., as in Nabavi et al., 2023b).

## 9. Challenges and Opportunities in MCDM

There are many challenges and opportunities in MCDM application; they are related to the criteria, alternatives, normalization, weights and the MCDM method for ranking the alternatives. All these can affect the ranking of alternatives and the recommended (usually, top-ranked) alternatives. In any application, all relevant criteria should be chosen. Then, all alternatives and the values of chosen criteria for each alternative will have to be found, which can be from suitable research and surveys, or by developing and solving the MOO problem with the chosen criteria as multiple objectives. Data on criteria values collected from research/surveys are likely to have variability and uncertainty, which can affect the ranking of the alternatives. On the other hand, solution of the MOO problem may not find all the Pareto-optimal solutions.

After having the relevant and reliable ACM, steps in MCDM often include normalization, finding criteria weights and use of one or more MCDM methods (Figure 2). Although certain normalization method, if required, is stated in the original paper of a MCDM method, it may or may not be the best. Many MCDM methods require criteria weights. So, one must choose one of many weighting methods or give weights for criteria; in either case, the best is unknown. Likewise, one must choose one or more methods from numerous MCDM methods available in the literature. The best MCDM method for an application or for one type of application is also not known. One main reason for many of these challenges is that the correct answer (i.e., top-ranked or recommended alternative) is unknown for a (chemical engineering) application. Additionally, MCDM methods may result in rank reversal phenomenon in some applications due to changes in ACM (Nabavi et al., 2023b).

The following strategies can be adopted to overcome (some of) the challenges mentioned above. One approach is to engage in deep and holistic thinking when selecting criteria. Another is to determine alternatives and their criteria values as reliably as possible, to ensure an accurate and comprehensive ACM. Additionally, trying different normalization, weighting, and/or MCDM methods for the application is recommended. Another effective strategy is to study the impact of changes in the ACM on the top-ranked alternative(s) identified by an MCDM method; essentially, perform sensitivity analysis on the top-ranked alternative(s) to potential variations in ACM, normalization, weighting, and the MCDM method. After reviewing the results from these tests, selecting the alternative recommended by most tests (i.e., majority voting) is advised. An effective computer program such as EMCDM, outlined in Section 6.1, would be highly conducive and convenient for conducting various tests in MCDM applications in chemical engineering and other disciplines.

With regards to sensitivity analysis, Wang et al. (2020) studied different normalization, weighting and MCDM methods for both mathematical and chemical engineering problems, and recommended entropy and CRITIC methods for weighting, and GRA, MABAC, SAW and TOPSIS. The subsequent study by Nabavi et al. (2023b) analyzed the effect of three types of changes in ACM on the top-ranked alternatives by eight popular/recent MCDM methods. They found that GRA, CODAS with entropy weights, and SAW with entropy or CRITIC weights are less sensitive to the studied changes in ACM. Baydaş et al.(2024) interpreted the determinants of sensitivity in MCDM using static reference rankings. Recently, Więckowski and Sałabun (2024) reviewed approaches for sensitivity analysis in MCDM in various disciplines.

Furthermore, the integration of MCDM with AI and ML represents a promising avenue for future research in chemical and process engineering. AI and ML techniques can significantly enhance MCDM applications by automating the processing of complex datasets, identifying non-linear relationships among criteria, and refining weighting methods by identifying underlying patterns in large ACMs. They can also enable advanced decision support systems by integrating predictive analytics with conventional MCDM models. This hybrid approach can be particularly beneficial for applications requiring fast, data-driven decisions such as real-time process monitoring in chemical plants. On the other hand, MCDM methods can be utilized to choose one of the alternative AI/ML models considering criteria such as accuracy, interpretability, computational cost, scalability, data requirements and generalizability. In short, future research can focus on developing frameworks that combine MCDM with ML algorithms like neural networks, decision trees, clustering, and dimensionality reduction techniques to process and analyze multi-dimensional ACMs and big data more efficiently.

## 10. Conclusions

In this paper, we reviewed the application of MCDM methods in chemical and process engineering, providing a comprehensive overview of the existing literature. We outlined selected studies to cover MCDM method, normalization technique, and weighting method employed, offering a detailed analysis of their applications in areas such as process optimization, sustainability assessment, and material selection. Readily available computer programs, e.g., EMCDM and PyMCDM, were outlined to highlight their utility in facilitating complex decision-making across different MCDM approaches, offering practitioners and researchers practical tools for handling multi-dimensional criteria.

Our review highlighted the progress made in MCDM applications within chemical and process engineering. Nonetheless, challenges remain, especially concerning the rank-reversal phenomenon, criteria weighting, and the stability of selected alternatives under varying conditions. Addressing these issues will require the development of adaptive frameworks and rigorous sensitivity analyses to ensure robust decision-making, regardless of changes in alternatives, criteria and/or weighting. The insights gained from this review contribute to a deeper understanding of the role of MCDM in advancing decision-making within chemical engineering, while also pointing promising directions for future studies. Integration of MCDM with AI and ML is one of the areas for further exploration, with potential applications in ML model selection, real-time process control, and predictive analytics. Future research should prioritize these hybrid frameworks, which could significantly enhance the applicability of MCDM, ultimately broadening the scope and impact of MCDM in chemical and process engineering.